\PassOptionsToPackage{numbers,sort&compress}{natbib}
\documentclass[preprint,12pt]{elsarticle}
\usepackage[margin=2.5cm]{geometry}
\usepackage{enumitem}
\newlist{inlinelist}{enumerate*}{1}
\setlist*[inlinelist,1]{%
  label=(\roman*),
}
\usepackage{soul}
\usepackage{hhline}
\usepackage{subfig}
\usepackage{multirow}
\usepackage{xcolor}
\usepackage{graphicx}
\usepackage{upgreek}
\usepackage{amssymb}
\usepackage{amsfonts,amsthm,bm,amsmath} 
\usepackage{appendix}
\usepackage{times}
\usepackage{caption}
\usepackage{subfig}

\usepackage{multirow}
\usepackage{xcolor}
\usepackage[linesnumbered,ruled,vlined]{algorithm2e}
\usepackage{tablefootnote}
\usepackage{comment}
\usepackage{bm}

\SetCommentSty{mycommfont}

\SetKwInput{KwInput}{Input}                
\SetKwInput{KwOutput}{Output}              

\usepackage[colorlinks]{hyperref} 
\hypersetup{ 
    colorlinks=true,       
    linkcolor=red,          
    citecolor=blue,        
    filecolor=magenta,      
    urlcolor=cyan           
}

\newcommand{\fref}[1]{Fig.~\ref{#1}}

\newcommand{\eref}[1]{Eq.~(\ref{#1})}

\newcommand{\sref}[1]{Section~\ref{#1}}
\newcommand{\tref}[1]{Table~\ref{#1}}
\setcitestyle{square}

\begin{document}

\begin{frontmatter}

\title{Advanced Deep Operator Networks to Predict Multiphysics Solution Fields in Materials Processing and Additive Manufacturing}
\author[]{Shashank Kushwaha$^{1}$ $^\ddagger$}
\author[]{Jaewan Park$^{1,2}$ $^\ddagger$}
\author[]{Seid Koric$^{1,2}$ 
\corref{mycorrespondingauthor}}
\cortext[mycorrespondingauthor]{Corresponding author}
\ead{koric@illinois.edu}
\author[]{Junyan He$^{1}$}
\author[]{Iwona Jasiuk$^1$}
\author[]{Diab Abueidda$^{2}$\corref{mycorrespondingauthor2}}
\cortext[mycorrespondingauthor2]{Corresponding author}
\ead{abueidd2@illinois.edu}
\address{$^1$ Department of Mechanical Science and Engineering, University of Illinois at Urbana-Champaign, Urbana, IL, USA \\
$^2$ National Center for Supercomputing Applications, University of Illinois at Urbana-Champaign, Urbana, IL, USA \\
}

\begin{abstract}
Unlike classical artificial neural networks, which require retraining for each new set of parametric inputs, the Deep Operator Network (DeepONet), a lately introduced deep learning framework, approximates linear and nonlinear solution operators by taking parametric functions (infinite-dimensional objects) as inputs and mapping them to complete solution fields. In this paper, two newly devised DeepONet formulations with sequential learning and Residual U-Net (ResUNet) architectures are trained for the first time to simultaneously predict complete thermal and mechanical solution fields under variable loading, loading histories, process parameters, and even variable geometries. Two real-world applications are demonstrated: 1- coupled thermo-mechanical analysis of steel continuous casting with multiple visco-plastic constitutive laws and 2- sequentially coupled direct energy deposition for additive manufacturing. Despite highly challenging spatially variable target stress distributions, DeepONets can infer reasonably accurate full-field temperature and stress solutions several orders of magnitude faster than traditional and highly optimized finite-element analysis (FEA), even when FEA simulations are run on the latest high-performance computing platforms. The proposed DeepONet model's ability to provide field predictions almost instantly for unseen input parameters opens the door for future preliminary evaluation and design optimization of these vital industrial processes.  
\end{abstract}


\end{frontmatter}
\def\thefootnote{$\ddagger$} \footnotetext{Both authors contributed equally to this work and should be considered as first authors}

\section{Introduction}
\label{sec:intro}

Continuous casting of steel and additive manufacturing are among the essential manufacturing processes for metal materials nowadays. Globally, almost 95\% of steel is produced through continuous casting. The severe environment of molten steel and the numerous process parameters that influence its intricate multiphysics phenomena restrict the number of experiments that may be conducted. The progress made in that well-established steel-making process mainly depends on the enhanced quantitative comprehension obtained from complex multiphysics numerical models used to optimize continuous casting and other steel-making processes and minimize their defects and $CO_2$ footprint. The World Steel Association reported that in 2022, the world's demand for steel grew 2.2\%,  and that the annual production of steel surpassed that of all other metals combined at 1881 million tons \cite{WorldSteelAssociation}. To put it in perspective, a small percent reduction in defects saves \$billions per year, including a significant reduction in $CO_2$ and other harmful emissions. 

On the other hand, metal additive manufacturing (AM) is a sophisticated technique for producing critical components layer by layer onto a part, guided by a digital 3D CAD model. Specifically, directed energy deposition (DED) is a metal AM technique that builds a component by melting metal feedstocks using focused thermal energy. DED can fabricate parts with high efficiency and flexibility for various applications, including fabrication of functionally-graded materials \citep{kelly2021directed} and repair damaged aero-engine turbine segments \citep{d2022design}. Further, Boeing 787s utilize various Titanium alloy components manufactured using DED technique to reduce production costs up to \$3 million \citep{froes2019additive}.  Although DED offers a lot of advantages over other classical and advanced manufacturing techniques, it is challenging to reduce cracks and porosity due to thermally induced residual stresses \citep{khanzadeh2019situ}. Hence, it is crucial to understand the intricate relationship between thermal distribution, residual stresses, and the process parameters, including printing speed and laser power, that can affect the part's mechanical properties. Like continuous casting, many complex multiphysics phenomena occur in the DED process at a high temperature in a short period of time, leading to high heating and cooling rates and sharp thermal gradients \citep{jardin2019thermal}. Such processes are difficult and expensive to measure through experiments, and numerical modeling is an effective tool for comprehending the fundamental physical mechanics of these processes \citep{song2020advances}.  


In recent years, extensive research studies have focused on improving the understanding of these processes using numerical simulations to analyze and enhance them. Steel solidification in continuous casting and metal additive manufacturing methods involve several physical phenomena across multiple temporal and spatial scales. Research has focused on microstructure material characterization, heat transfer modeling, phase change kinetics including melting and solidification, residual stress, buoyancy, and fluid flow analysis. Various length scales in additive manufacturing give rise to different modeling approaches: microscopic, mesoscopic, and macroscopic scales. Microscopic models \cite{rodgers2021simulation, johnson2018simulation, korner2020modeling} are primarily utilized to study the microstructure development resulting from the rapid melting and solidification process. Mesoscopic models \cite{khairallah2014mesoscopic, king2015laser,zhang2023additive, tang2018numerical} examine how lasers interact with individual powder particles and are utilized to analyze problems in additive manufacturing, like porosity and surface roughness. Large-scale continuum macroscopic models \cite{koric2006efficient, koric2010multiphysics, hodge2014implementation, parry2016understanding, denlinger2017thermomechanical,AHMED2023e19385} aim to simulate the evolution of product-scale processes by simplifying assumptions to ensure computational efficiency. These assumptions include modeling powder in additive manufacturing or mushy and liquid zones in steel casting as an effective continuum and disregarding fluid flow and other sub-continuum effects. The thermomechanical macro models predict the temperature field and estimate residual stress and distortion. Various multi-scale modeling frameworks were designed to incorporate information from the micro and mesoscale models back into macro-scale models to provide even more detailed and accurate predictions. Interested readers can refer to \cite{bayat2021review} for a review of the multi-scale models. Developments in simulating the continuous casting and DED have also been made by employing computational fluid dynamics (CFD) to predict rapidly evolving geometries that are accompanied by significant mass and heat transfer during the flow of molten material. Still, they cannot predict the development of residual stress. A fully multiphysics thermo-fluid-mechanical modeling framework that couples and exploits the benefits of both techniques while avoiding their respective limitations was also developed \cite{koric2010enhanced, zappulla2020multiphysics, mathews2023temporally}. 

The rapid developments in computational hardware and algorithms in the last ten years allow these simulations to be performed on high-performance compute clusters. However, computational complexities from high-fidelity meshes discretizing large domains on the macro-level to model multi-physics phenomena in these processes accurately prevented designs, optimizations, sensitivity analysis, uncertainty quantifications, online controls, and other computationally challenging workflows that require a large number of instant forward evaluations \cite{keyes2013multiphysics, bayat2023holistic}. At the same time, machine learning (ML) has progressed quickly in recent years, leading to various applications in fields like image and speech recognition, medical diagnosis, autonomous driving, bioinformatics, bio-inspired design, and many more \citep{noda2015audio, grigorescu2020survey, kushwaha2023designing}. Artificial neural networks, a type of machine learning based on the structure and function of the brain, have gained significant attention in physics-based modeling \cite{tu2023integrating, sun2023physics} to generate surrogate models for complex physical phenomena. A well-trained surrogate deep learning model can quickly infer results comparable to traditional modeling methods without requiring high-performance computer resources or specialized modeling tools. Deep learning methods have been trained and used in various practical manufacturing assembly processes, such as by data collected from sensors \cite{shao2020ieee} or data generated by classical numerical methods \cite{SHAHANE2022106843}. Moreover, they have also been applied in AM and continuous casting \cite{meng2020machine, jin2020machine, koric2021deep, perumal2023temporal, kats2022physics, met_am_overview_2021}.

Choi et al. \cite{choi2024accelerating} adopted 3D U-Net architecture in laser power bed fusion to build a surrogate model predicting microstructure evolution. However, conventional surrogate neural network models typically predict solutions at particular locations. If they manage to predict the solution for a specific field with a particular set of parameters, even a tiny alteration in input parameters like loads, boundary conditions, or domain geometry necessitates expensive re-training or transfer learning.  Lu et al. \cite{lu2021learning} created a DeepONet structure to tackle this important problem. DeepONet can effectively map an infinite parameter space to the complete solution field on the computing grid because of its dual network structure consisting of trunk and branch networks. A well-trained DeepONet may quickly infer a complete solution field for a fresh set of parameters without requiring retraining or transfer learning, significantly outperforming standard numerical methods in speed.   DeepONets have been utilized to forecast entire solution fields in several engineering and scientific domains, such as materially nonlinear solid mechanics \cite{koric2023deep}, fracture \cite{goswami2022physics}, aerodynamics \cite{zhao2023learning}, acoustics \cite{xu2023training}, seismology \cite{haghighat2024deeponet}, digital twin \cite{kobayashi2024improved} and heat transfer \cite{sahin2024deep, koric2023data}. The original DeepONet utilized forward fully connected subnets. Recently, new formulations have been developed, including the sequential DeepONet \cite{he2024sequential} designed to handle time-dependent inputs and the DeepONet using ResUNet in its trunk \cite{HE2023116277} to address issues related to variable 2D input geometries under parametric loads and elastoplastic material behavior. Finally, Yaseen et al. \cite{yaseen2023fast} benchmarked the performance of DeepONet, Fourier neural operator (FNO) \cite{li2020fourier}, and classical deep neural network (DNN) to predict temperature field in a single physics thermal model of direct energy deposition (DED) on a simple geometry. It was found that operator models outperform the DNN model in terms of accuracy and generalizability. 

In this novel work, it is the first time, best to the author's knowledge, that the advanced neural network operator architectures \cite{he2024sequential, HE2023116277} were trained with high-fidelity data from multiphysics simulations in continuous casting and DED AM process to simultaneously and relatively accurately predict entire temperature and stress fields for unknown input parameters representing variable loads and geometries in AM and real-world time histories of thermal and mechanical loads in casting, thus, opening the door for preliminary design, optimizations, online controls, and other exceptionally computationally demanding or even currently unsolvable modeling and analysis challenges in these critical industrial processes.  

The structure of this paper is as follows: \sref{sec:methods} describes the data generation process and offers a comprehensive explanation of the neural networks employed in this study. \sref{sec:results} presents the performance of the neural networks and interprets their significance, including the stress prediction application in AM with trained ResUNet-based DeepONet.  Lastly, \sref{sec:conc} concludes by summarizing the findings and proposing directions for future research.

\section{Methods}
\label{sec:methods}
\subsection{Training data generation}
\label{sec:data_gen}
\subsubsection{Thermo-Mechanical Model of Steel Solidification in Continuous Casting}
\label{sec:data_solidification}
In the first use case, we utilize our Sequential DeepONet (S-DeepONet) to predict the final temperature and stress fields in a multiphysics simulation of solidifying steel during the continuous casting process. To create training and testing data for S-DeepONet, we leverage the modeling results from the existing multiphysics (thermo-mechanical) model \cite{koric2006efficient}, which has a solidifying slice traveling in the Lagrangian frame of reference down the continuous caster.  The governing equation for the solidifying shell's thermal behavior is expressed as:
\begin{equation}
    \rho( \frac{\partial H}{\partial t} )= \nabla \cdot ( k\nabla T )
    \label{solidifying_GDE}
\end{equation}
where $\rho$ is density, $k$ is thermal conductivity, and $H$ is specific enthalpy which includes the latent heat during phase transformations, such in transition from liquid to solid or in solid phases transitioning from delta-ferrite to austenite. 
Since inertia effects are negligible in solidification, the quasi-static mechanical equilibrium governs the mechanical behavior is written as: 
\begin{equation}
    \boldsymbol{\nabla}\cdot\boldsymbol{\sigma}(\textbf{X})+\textbf{b}=0
    \label{solidifying_Equilibrium}
\end{equation}
where $\boldsymbol{\sigma}$ is the Cauchy stress tensor and \textbf{b} is the body force density vector. The rate representation of total strain can be split into elastic strain rate $\dot{\boldsymbol{\varepsilon}}_{el}$, inelastic strain rate (visco-plastic) $\dot{\boldsymbol{\varepsilon}}_{ie}$, and thermal strain rate $\dot{\boldsymbol{\varepsilon}}_{th}$:
\begin{equation}
    \dot{\boldsymbol{\varepsilon}}=\dot{\boldsymbol{\varepsilon}}_{el} + \dot{\boldsymbol{\varepsilon}}_{ie} + \dot{\boldsymbol{\varepsilon}}_{th}.
    \label{total_strain}
\end{equation}

The stress and strain rates are related by the constitutive equation:
\begin{equation}
    \dot{\boldsymbol{\sigma}}=\textbf{D}:(\dot{\boldsymbol{\varepsilon}}-\alpha\textbf{I}\dot{T}-\dot{\boldsymbol{\varepsilon}}_{ie}),
    \label{solidifying_Constitutive}
\end{equation}
where \textbf{D} is the fourth-order tensor containing the isotropic elasticity constants: the temperature-dependent elastic modulus and Poisson ratio. $\alpha$ is the thermal expansion coefficient tensor, and \textbf{I} is the identity tensor. 
Steel alloys exhibit notable temperature and time-dependent plastic behavior, including phase transformations at high temperatures, particularly around the solidification/melting point. To represent the austenite phase of steel, Kozlowski et al. \cite{kozlowski1992simple} developed a visco-plastic constitutive equation that connected inelastic strain to temperature, strain rate, stress, and steel grade through carbon content: 
\begin{equation}
    \begin{aligned}
        &\dot{\bar{\boldsymbol{\varepsilon}}}_{ie}[sec^{-1}] = f_{c}(\bar{\boldsymbol{\sigma}}[MPa]-f_{1}\bar{\boldsymbol{\varepsilon}}_{ie}|\bar{\boldsymbol{\varepsilon}}_{ie}|^{f_{2}-1})^{f_{3}}exp(- \frac{Q}{T[K]}) \\
        &\text{where:} \\
        &Q = 44,465 \\
        &f_{1} = 130.5 - 5.128 \times 10^{-3}T[K] \\
        &f_{2} = -0.6289 + 1.114 \times 10^{-3}T[K] \\
        &f_{3} = 8.132 - 1.54 \times 10^{-3}T[K] \\
        &f_{C} = 46,550 - 71,400(\%C) + 12,000(\%C)^{2}
    \end{aligned}
    \label{visco_plastic_Constitutive}
\end{equation}
where $\boldsymbol{\sigma}$ (MPa) is von Mises effective stress, $Q$ is an activation energy constant, \%C is carbon content (weight percent) representing steel grade (composition), and $f_{1}, f_{2}, f_{3}, f_{4}$ are empirical functions that depend on absolute temperature ($K$). \\
In order to model the delta-ferrite phase, which is weaker than the austenite phase and has a substantially higher creep rate, a different constitutive model known as the Zhu power law model \cite{zhu1996coupled} was developed:
\begin{equation}
    \begin{aligned}
        &\dot{\bar{\boldsymbol{\varepsilon}}}_{ie}(1/sec.) = 0.1 \left |\frac{\bar{\boldsymbol{\sigma}}(MPa)}{f_{\delta c}(\%C) \left (\frac{T(^{\circ}K)}{300} \right ) ^{-5.52}(1+1000\bar{\boldsymbol{\varepsilon}}_{ie})^{m}} \right |^{n} \\
        &\text{where: } \\
        &f_{\delta c}(\%C) = 1.3678 \times 10^{4} (\%C)^{-5.56 \times 10^{-2}} \\
        &m = -9.4156 \times 10^{-5} T(^{\circ}K) + 0.3495 \\
        &n = \frac{1}{1.617 \times 10^{-4}T(^{\circ}K) - 0.06166}.
    \end{aligned}
    \label{zhu_power_law}
\end{equation}
When the volume fraction of delta-ferrite in a solid exceeds 10\%, the delta-phase model in \eref{zhu_power_law} is employed to simulate the dominant effect of the extremely high creep rates in the delta-ferrite phase of mixed-phase structures on the net mechanical behavior. An elastic-perfectly-plastic constitutive model with small yield stress is applied in mushy and liquid zones (above the solidus temperature Tsol) to enforce negligible strength in those volatile zones.

\begin{figure}[h!] 
    \centering
         \includegraphics[width=0.4\textwidth]{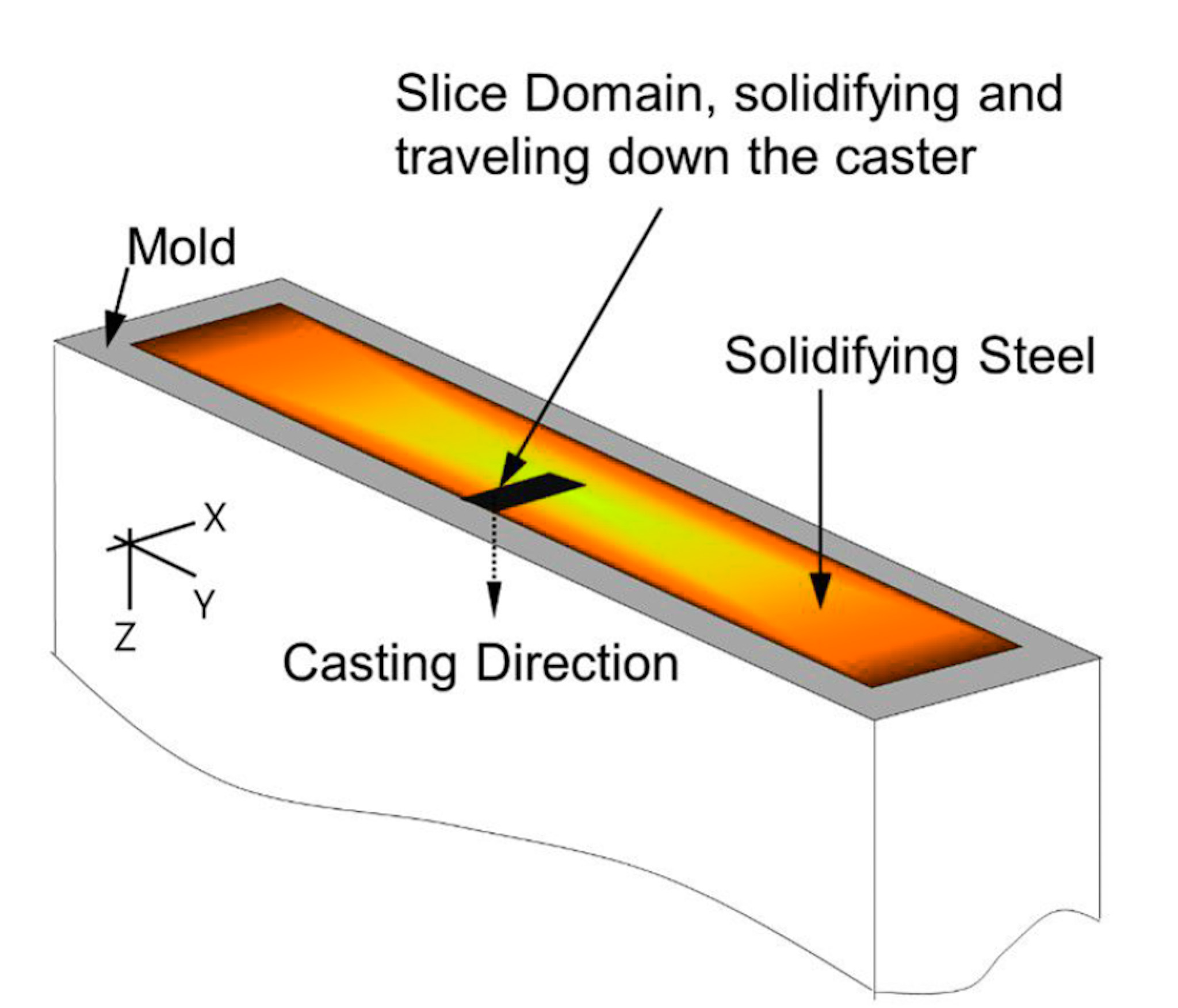}
    \caption{Continuous caster with solidifying slice finite element domain}
    \label{Slice}
\end{figure}

The thermal strain term in \eref{solidifying_Constitutive} depends on the isotropic temperature-dependent thermal expansion coefficient $\alpha$, as well as evolving temperature distribution. Other material properties, such as conductivity, density, enthalpy, and elastic modulus, as well as the constitutive models, are also strongly dependent on temperature, while temperature depends on mechanical deformation and thermal contraction of the solidifying shell through interfacial heat transfer extracting heat to the mold. All this calls for a strongly coupled solution strategy. The highly nonlinear constitutive visco-plastic models in \eref{visco_plastic_Constitutive} and \eref{zhu_power_law} are integrated by a special bounded Newton-Raphson method at the integration points in UMAT subroutine \cite{koric2006efficient} and linked with Abaqus software \cite{Abaqus2022}, which solves the coupled thermo-mechanical governing  \eref{solidifying_GDE} and \eref{solidifying_Equilibrium} by the implicit nonlinear finite element solution methodology.  The phase fraction and temperature-dependent material properties calculations for the low carbon steel grade chosen for this work with 0.09 wt\%C, with solidus and liquids temperatures $T_{sol}$ = 1480.0 $^{\circ}C$, and $T_{liq}$ = 1520.7 $^{\circ}C$ are also an integral part of the UMAT subroutine, and more details about this robust multiphysics model can be found in \cite{zappulla2020multiphysics}. When subjected to thermal loading, generalized plane strain conditions can effectively restore a complete three-dimensional stress state in long objects \cite{koric2006efficient, koric2010multiphysics, zappulla2020multiphysics}, such as the continuous caster in the left side of \fref{Slice} with its significant length and width. In a Lagrangian frame of reference, the slice domain in \fref{Slice} moves down the mold at casting velocity.


\begin{figure}[h!] 
    \centering
         \includegraphics[width=0.85\textwidth]{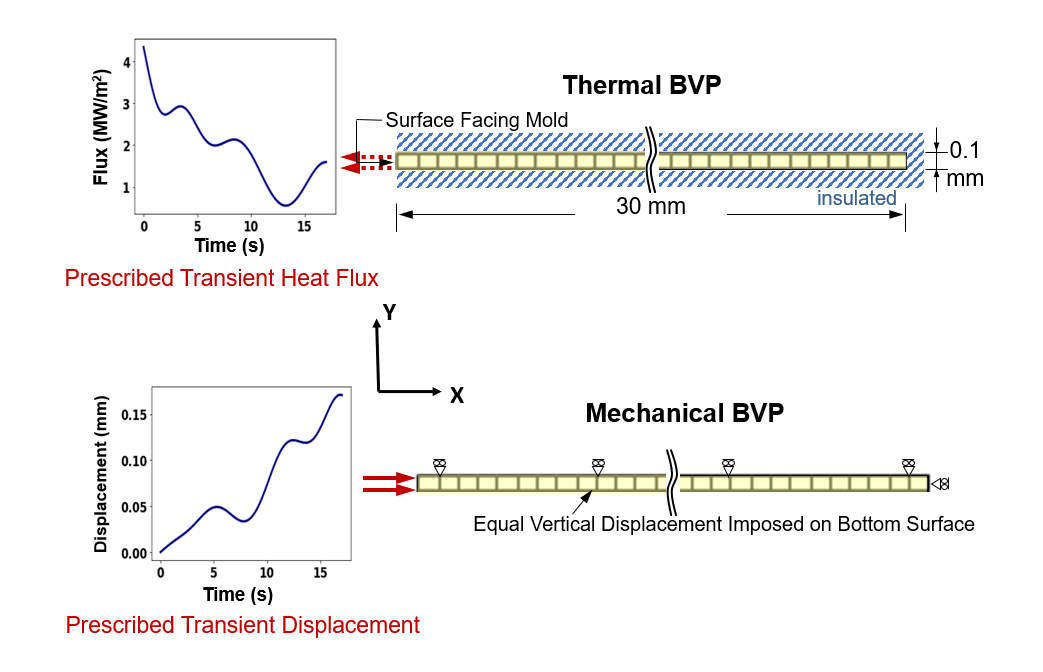}
    \caption{Thermal and Mechanical BVPs with Random Input Profiles}
    \label{Slice_BVP}
\end{figure}

\fref{Slice_BVP} shows the thermal and mechanical conditions used in solving its corresponding boundary value problems (BVP) to generate training and testing data. The generalized plane strain condition is provided in the axial (z-direction) direction by a single row of 300 connected thermo-mechanical generalized plane strain elements with 602 nodes. Furthermore, a second generalized plane strain condition was enforced along the domain's bottom border by imposing equal strain in the y direction on all nodes' vertical displacements. The thermal and mechanical boundary conditions of the solidifying slice domain are provided by time-dependent profiles of thermal fluxes leaving the chilled surface on the left side of the domain and their displacement due to mold taper and other interactions with the mold. 
The studies made at the steel plant indicate that while the displacement profile shows an increasing trend because of mold taper, the thermal flux has an overall decreasing profile because of transient heat transfer. A radial basis interpolation with the Gaussian function is used to connect (interpolate) the several temporal points that have been randomly defined within ranges of expected profile values enabling additional fluctuations and data noise that are observed in the actual flux and displacement profiles due to abrupt changes in contact conditions and interfacial heat transfer between the mold and steel surfaces. The calculated temperature and stress fields are the targeted outputs of the S-DeepONet method, while the temporal fluctuations of imposed thermal flux and displacement profiles on the chilled surface of the solidifying shell serve as inputs. A total of 10,000 data points were produced with these thermal flux and displacement input profiles, offering every circumstance in which the solidifying shell would meet on its chilled surface while moving down the caster.

Chronic defects affect the continuous casting and other steel-making processes, including cracking, surface depression, and hot tearing problems. They primarily develop during the initial solidification in the mold and, to be adequately modeled and predicted, require full temperature and lateral ($\sigma_{22}$) component stress solutions. However, simultaneously learning and predicting full-field temperature and stress results with deep learning methods along the solidification slice based on two independent input profiles in this highly coupled thermo-mechanical simulation is challenging. This is particularly the case for the lateral ($\sigma_{22}$) stress solution component, which is highly sensitive to phase transformations that happen during solidification between liquid/mushy, austenite, and delta-ferrite phases – each driven by a separate constitutive law, causing non-smooth and noisy stress solution profiles with sharp spikes.    

\subsubsection{Thermo-Mechanical Model of Additive Manufacturing}
\label{sec:AM_data_generation}

\begin{figure}[h!] 
    \centering
     \subfloat{
         \includegraphics[trim={0cm 0cm 0cm 0cm},clip,width=0.25\textwidth]{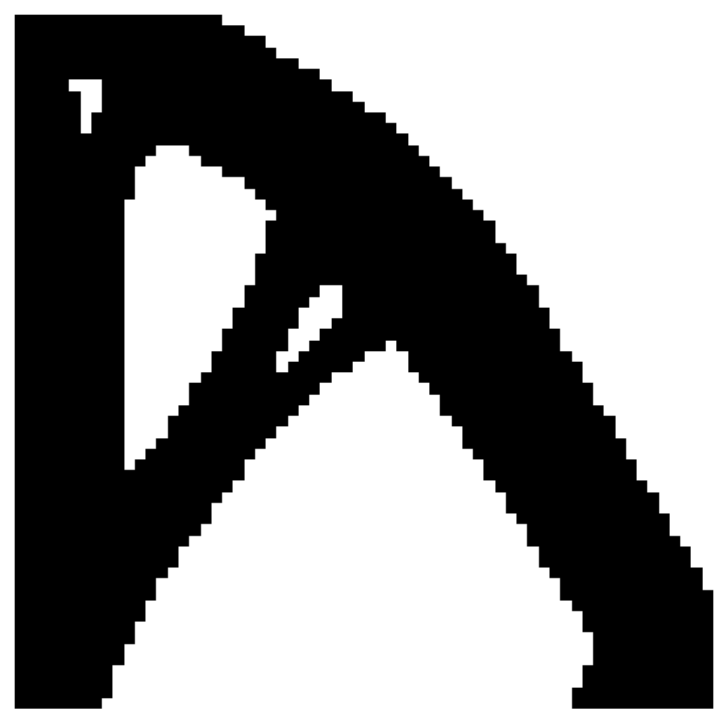}
         \label{d0}
     }
     \subfloat{
         \includegraphics[trim={0cm 0cm 0cm 0cm},clip,width=0.25\textwidth]{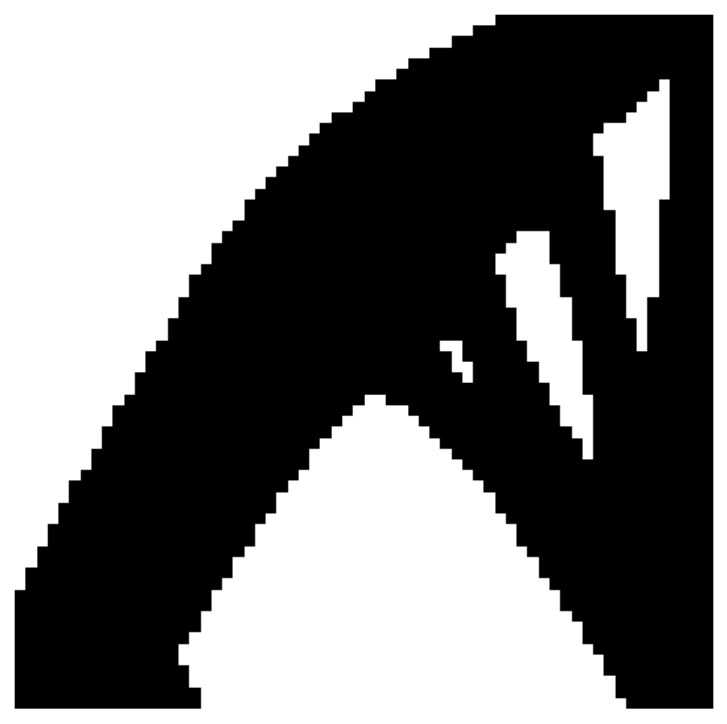}
         \label{d1}
     }
     \subfloat{
         \includegraphics[trim={0cm 0cm 0cm 0cm},clip,width=0.25\textwidth]{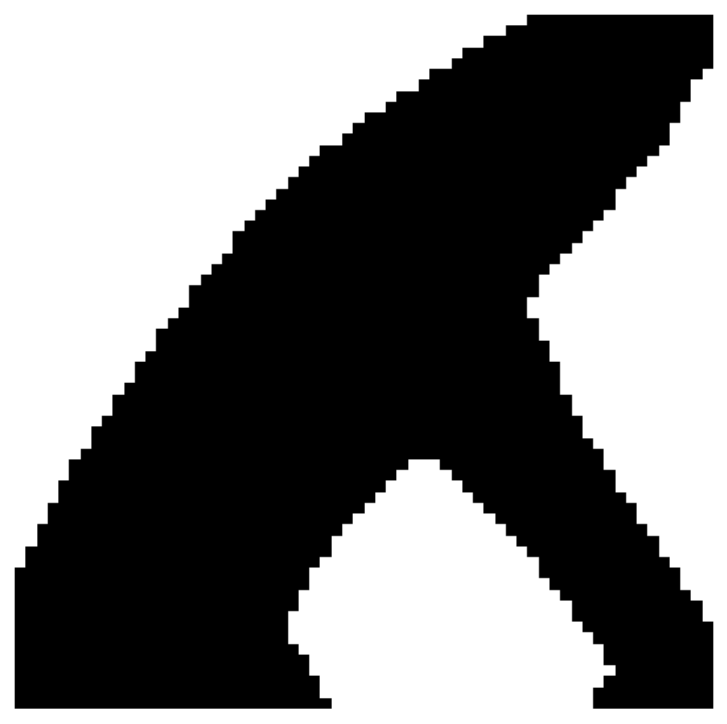}
         \label{d2}
     }
    \caption{Three typical designs from topology optimization.}
    \label{to_designs}
\end{figure}

In the second use case, ResUNet-based DeepONet \cite{HE2023116277} is exploited to forecast the final temperature and residual stress fields in the LDED AM process. Unlike the first use case, this problem considers a sequentially coupled thermo-mechanical multiphysics model. First, we perform the transient heat transfer problem by applying the thermal loads to the printed design during the deposition process. Following this, a quasi-static structural analysis is conducted, driven by the temperature field obtained from the thermal analysis. During the LDED process, the material is deposited by a nozzle mounted on a multi-axis arm and simultaneously melted by a laser heat source. The new material solidifies sequentially in a layer-by-layer fashion until the 3D designs are built. In this example, designs generated from a compliance-minimizing elastic topology optimization are used (for details on the topology optimization problem, the readers are referred to \cite{HE2023116277}). Selected topology-optimized designs are shown in \fref{to_designs} with different volume fractions. We assume these designs will be printed vertically with no supports via LDED. The 3D printing process is simulated via a thermal-mechanical analysis plugin, AM Modeler \citep{AM_Modeler}, part of Abaqus/Standard \citep{Abaqus2022}. This plugin has been successfully used in the literature for the LDED process \citep{song2020advances}. The plugin eliminates the need to use various user-defined subroutines. It can be accessed through table collections with string identifiers beginning with "ABQ\_AM," allowing users better control over the process parameter definition. The mesh elements are progressively activated using full element activation during printing. The cross-section of the deposited material is assumed to be rectangular, with a bead width of 3.35 mm and a bead height of 1 mm.  

\begin{figure}[h!] 
    \centering
         \includegraphics[width=0.35\textwidth]{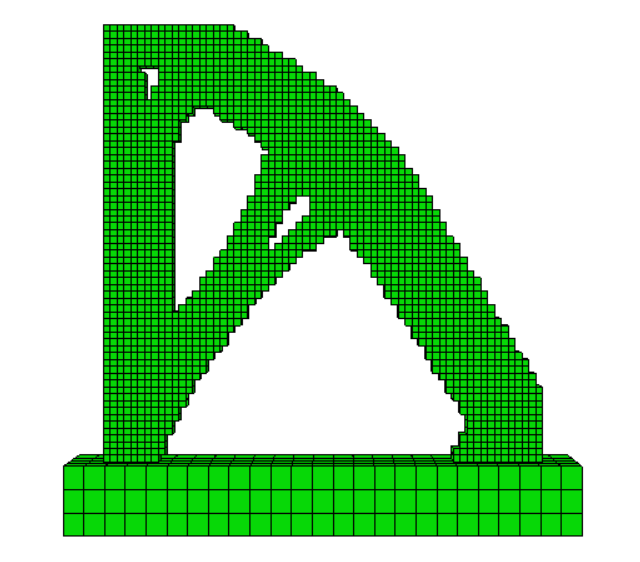}
    \caption{A typical mesh used in thermal-mechanical simulation.}
    \label{AM_mesh}
\end{figure}

The mesh used for a representative design in the finite element model is shown in \fref{AM_mesh}. The structure is modeled with uniform mesh using 8-node tri-linear brick elements. The heat transfer analysis and static structural analysis share the same mesh strategy. The heat transfer analysis uses DC3D8 elements, while structural analysis uses C3D8 elements. The element dimensions for the printed structure are 3.35mm x 1mm x 1mm. A coarser mesh is used for the supporting substrate. The TO design and the substrate are made of material Inconel 625. The temperature-dependent properties, including thermal conductivity, specific heat, the coefficient of thermal expansion (CTE), elastic modulus, and yield stress, are summarized in \tref{temp_dependent_prop} \citep{Inconel_properties}. The density is set to 8440 kg/mm$^3$. The solidus temperature is 1290$^{o}$C, the liquidus temperature is 1350$^{o}$C, and the latent heat of fusion is 2.72 x 10$^{5}$ J/kg. The Poisson's ratio is 0.366 \citep{denlinger2015effect}. 

\begin{table}[h!]
    \caption{Temperature-dependent material properties for Inconel 625 \citep{Inconel_properties}}
    \small
    \centering
    \begin{tabular}{c|ccccc}
    
     \multicolumn{1}{c|}{\textbf{Temperature}} & \multicolumn{1}{c}{\textbf{Thermal}} & \multicolumn{1}{c}{\textbf{Specific}} & \multicolumn{1}{c}{\textbf{CTE}} & \multicolumn{1}{c}{\textbf{Elastic}} & \multicolumn{1}{c}{\textbf{Yield}} \\
     \multicolumn{1}{c|}{($^\circ$C)} & \multicolumn{1}{c}{\textbf{Conductivity}} & \multicolumn{1}{c}{\textbf{Heat}} & \multicolumn{1}{c}{($1/^\circ$C)} & \multicolumn{1}{c}{\textbf{Modulus}}  & \multicolumn{1}{c}{\textbf{Stress}} \\
     & \multicolumn{1}{c}{(mW/(mm·$^\circ$C))} & \multicolumn{1}{c}{(mJ/(tonne·$^\circ$C))} & \multicolumn{1}{c}{} & \multicolumn{1}{c}{(MPa)} & \multicolumn{1}{c}{(MPa)} \\
      & & x ($10^8$) & x ($10^{-5}$) & x ($10^{5}$)  & \\
    \hline
    20 & 9.9  & 4.10 & 1.28 & 2.08  & 493 \\
    93 & 10.8 & 4.27 & 1.28 & 2.04  & 479 \\
    205 & 12.5 & 4.56 & 1.31 & 1.98  & 443 \\
    315 & 14.1 & 4.81 & 1.33 & 1.92  & 430 \\
    425 & 15.7 & 5.11 & 1.37 & 1.86  & 424 \\
    540 & 17.5 & 5.36 & 1.40 & 1.79  & 423 \\
    650 & 19.0 & 5.65 & 1.48 & 1.70  & 422 \\
    760 & 20.8 & 5.90 & 1.53 & 1.61  & 415 \\
    870 & 22.8 & 6.20 & 1.58 & 1.48  & 386 \\
    
    \end{tabular}
    \label{temp_dependent_prop}
\end{table}

The transient heat transfer problem in the LDED process is solved by utilizing a 3D heat conduction equation. The temperature field is determined by solving heat conduction equation with given initial conditions, heat source, and surface boundary conditions. The equation for 3D heat conduction is given by: 
\begin{equation}
\label{eq_heat_conduction}
    \rho(T) C_{p}(T)( \frac{\partial T}{\partial t} )= \nabla \cdot ( k\nabla T ) + q(r, t),
\end{equation}
where $\rho$, $T$ $C_{p}$, $k$, $q$, and  $t$ are the density [kg/m$^3$], temperature [K], specific heat [J/ kg K], thermal conductivity [W/m K], volumetric heat source [W/m$^3$], and time[s], respectively. The variable $r$ can be replaced by each coordinate (i.e., $x, y, z$ in the Cartesian coordinates). The initial temperature ($T_{0}$) at $t = 0$ was set to 26$^o$C. Further, a portion of heat is dissipated by convection and radiation. Due to new material deposition, previously exposed surfaces are covered, and new free surfaces are generated. Surface convection and radiation are defined on the continuously evolving free surfaces. The surface convection and radiation heat loss during the printing process are defined by:
\begin{equation}
\label{eq_heat_convection}
    q_{convection} = h (T - T_{\infty})
\end{equation}
\begin{equation}
\label{eq_heat_radiation}
    q_{radiation}= \tilde{\epsilon} k_{B}(T^{4} - T^{4}_{\infty})
\end{equation}
where $h$ is the convection heat transfer coefficient set to 18 W/m$^2$ $^{o}$C, $T_{\infty}$ is the ambient chamber temperature set to 26$^o$C, and $k_{B}$ is the Stefan-Boltzmann constant. The emissivity ($\tilde{\epsilon}$) is 0.28. The temperature fields from heat transfer analysis were used as inputs for the structural analysis to calculate the thermal and total strains, as follows:
\begin{equation}
\label{eq_thermal_strain}
    \varepsilon_{th}= \alpha (T - T_{0})
\end{equation}
\begin{equation}
\label{eq_total_strain}
    \boldsymbol{\varepsilon}= \boldsymbol{\varepsilon_{e}} + \boldsymbol{\varepsilon_{pl}} + \boldsymbol{\varepsilon_{th}}
\end{equation}
where $\alpha$ is the thermal expansion coefficient. The total, elastic, plastic, and thermal strains are given by $\boldsymbol{\varepsilon}$, $\boldsymbol{\varepsilon_{e}}$, $\boldsymbol{\varepsilon_{pl}}$, and $\boldsymbol{\varepsilon_{th}}$, respectively. The total strain is utilized for the elastic perfectly-plastic mechanical analysis to calculate residual stress in the LDED process:
\begin{equation}
\label{eq_ss}
    \boldsymbol{\sigma}= \textbf{D} : \boldsymbol{\varepsilon_{e}}
\end{equation}

\begin{equation}
\label{eq_stiffness}
    D_{ijkl}= \lambda \delta_{ij} \delta_{kl} + \mu (\delta_{ik} \delta_{jl} +\delta_{il} \delta_{jk})
\end{equation}

\begin{equation}
\label{eq_stiffness}
    \lambda= \frac{E \nu}{(1+ \nu)(1 - 2\nu)}
\end{equation}

\begin{equation}
\label{eq_stiffness}
    \mu= \frac{E}{2(1+\nu)}
\end{equation}

\begin{equation}
\label{eq_delta}
  \delta_{ij} =
    \begin{cases}
      1, & \text{$i=j$}\\
      0, & \text{$i\neq j$}
    \end{cases}       
\end{equation}
where $\lambda$ and $\mu$ are known as Lam\'e's constant. $E$ and $\nu$ are the Young's modulus and Poisson's ratio, respectively. Also, \eref{eq_stiffness} and \eref{eq_delta} use indicial notation to present fourth-order elastic stiffness tensor (\textbf{D}) for isotropic material and Kronecker delta ($\delta$) function, respectively. For simplicity, the flow stress of the perfectly plastic material after initial yield is dependent on local temperature but not on the local design feature size, effectively ignoring the apparent size-dependent material properties commonly observed when printing thin features \cite{he2023size,june2024effects}. The temperature-dependent yield stress is shown in \tref{temp_dependent_prop}, where

\begin{equation}
\label{yield_stress}
    \overline{\sigma} (\boldsymbol{\varepsilon_{pl}}) = \sigma_{Y} (T).
\end{equation}
In this work, the von Mises yield criterion presented in \eref{yield_criterion} is used. 
\begin{equation}
\label{yield_criterion}
    \Bar{\sigma} = \sqrt{ \frac{(\sigma_{11} - \sigma_{22})^2 + (\sigma_{22} - \sigma_{33})^2 + (\sigma_{33} - \sigma_{11})^2 + 6(\sigma_{12}^2 + \sigma_{23}^2 + \sigma_{31}^2)}{2} }.
\end{equation}

An accurate heat input model is required to predict the nodal temperatures and temperature fields. The Goldak heat source model has been used widely for welding and AM processes \citep{goldak1984new, yang2019residual}. Hence, this work uses a moving heat flux with a Goldak distribution to simulate the laser-induced heating during the deposition. The laser beam spot at the part surface is assumed to be circular. The energy absorption efficiency is taken to be 0.4 for all the cases. In the static structural analysis, the initial temperature of the deposited material is assigned as the solidus temperature, 1290$^o$C, while the initial substrate temperature remains at the room temperature, 26$^o$C. During the deposition process, the beads in a layer are deposited in the same direction, and the direction alternates between the layers. The deposition process is simulated in 64 layers ($L_{total}$). The raw material is melted upon deposition by a laser power of 3 kW. The process parameters are taken from the previous experimental study \citep{denlinger2015effect} and the example problem defined in Abaqus AM user manual \citep{AM_Modeler_LDED_example}.  

A 1000 topology-optimized designs are considered for the analysis. Each simulation is repeated for 5 different print speeds [7.5, 10.0, 12.5, 15.0, 17.5] mm/s. The total time ($t_{total}$) for all the simulations is kept constant at 685 s. The print time ($t_{print}$) is calculated using the design length (64 mm) and the corresponding nozzle speed (mm/s). Then cooling time ($t_{cooling}$) can be calculated using:
\begin{equation}
\label{eq_cooling_time}
    t_{cooling} = \left(\frac{t_{total}}{L_{total}}  \right) - t_{print}.
\end{equation}
An event series file defines the material deposition sequence, the laser scanning path and laser power while printing and cooling at the current position. 

For heat transfer analysis, nodal temperature field output is requested for the whole model at specified time increments. For static structural analysis, the temperature field stored in the output database from the heat transfer analysis is used as input in the predefined field. Further, nodal displacement and stress field output are requested for the whole model. After the data extraction, 4574 valid simulations were used for deep learning with the temperature fields and residual stress for the last time step in the deposition process.

The quality and performance of AM designs under various mechanical loadings and applications are significantly influenced by residual stresses. These stresses, in turn, are closely linked to the corresponding temperature field. This relationship complicates the concurrent prediction of temperature and stress fields. Additionally, 5,000 intricate shapes resulting from the topology optimization process further add to the challenge makes it heavily compound. Consequently, in our approach, we select ResUNet-based DeepOet to enhance the model's predictive accuracy, incorporating the printing velocity as an input parameter and utilizing the extracted nodal temperature and stress fields as targets.


\subsection{Neural network models}
\label{sec:NNs}
Two distinct neural networks were utilized for each dataset to accurately predict von Mises stress and temperature across the entire field. The DeepXDE platform developed by Lu et al. \cite{lu2021deepxde} was employed, using a Tensorflow backend. 


\subsubsection{S-DeepONet}
\label{sec:s-deeponet}
\begin{figure}[h!] 
    \centering
    \includegraphics[width=0.95\textwidth]{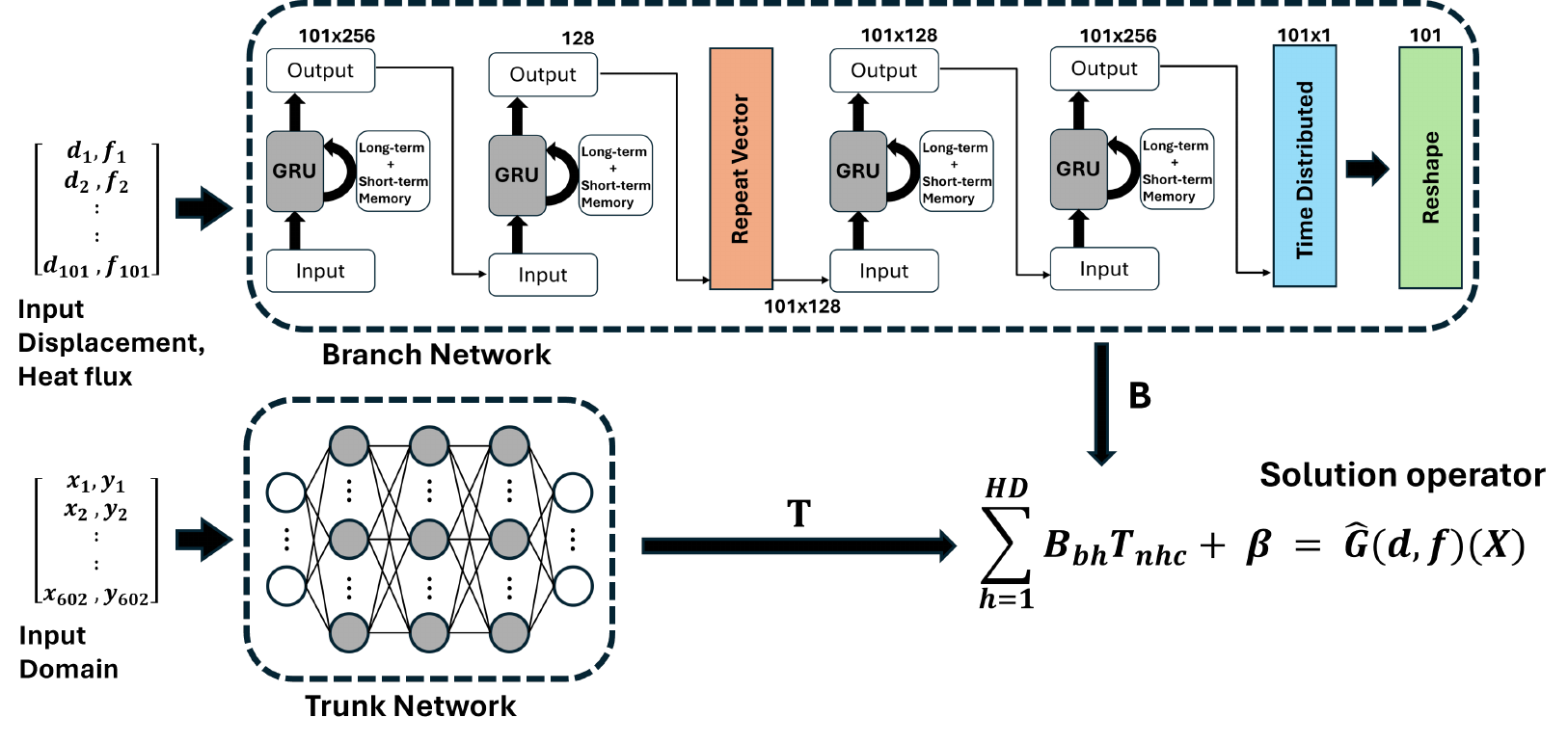}
    \caption{Architecture of the multi-component S-DeepONet for multiphysics problems. $d$ and $f$ represent the time-dependent input displacement and heat flux used in simulation, and $x$, $y$ are nodal coordinates. $\Hat{G}$, as a solution operator, produces final von Mises stress and temperature for each node in the domain.}
\label{architecture}
\end{figure}

S-DeepONet, an extended concept of DeepONet compatible with time-dependent input data, was recently introduced by the authors \cite{he2024sequential}. A gated recurrent unit (GRU) model in the branch network processes time-dependent sources, and a feed-forward neural network (FNN) model in the trunk network handles the spatial input of differential equations. In this work, we used this model on the data generated in \ref{sec:data_solidification}. The network was extended to accept multiphysics inputs (applied displacement and heat flux) and predict 2 solution components (stress and temperature). The encoded outputs of the branch and trunk networks are combined together via a matrix-vector product:

\begin{equation}
    \bm{G}_{bnc} = \sum_{h=1}^{HD} \bm{B}_{bh} \bm{T}_{nhc} + \boldsymbol{\beta}
    \label{combine}
\end{equation}
where $\bm{G}$ represents the solution operator of S-DeepONet, while $\bm{B}$ and $\bm{T}$ corresponds to the outputs of the branch and trunk networks, respectively. The hidden dimension (HD) is the latent space size. Meanwhile, $n$ denotes the number of points in the domain, and $c=2$ is the number of output components. Lastly, $\boldsymbol{\beta}$ represents the bias incorporated into the product. This framework enables the prediction of comprehensive field solutions and can forecast multiple solution fields in complex multi-physics problems.
The multi-component version of the S-DeepONet structure is illustrated in \fref{architecture}.

As illustrated in \fref{architecture}, the initial GRU cell, equipped with 256 units, produces an output in the form of a $101 \times 256$ matrix. Following this, the subsequent GRU cell, configured to not return sequences, takes this matrix as input, producing a vector of 128 dimensions in length. A repeat vector cell is then employed to replicate this vector into a matrix of dimensions $101 \times 128$. The third GRU cell maintains this matrix dimensionality, $101 \times 128$. The final GRU cell processes this matrix to output a new matrix of dimensions $101 \times 256$. A Time Distributed Dense layer is applied to the output of the GRU, resulting in a vector of dimensions $101 \times 1$, which is then reshaped into a vector of length 101, effectively eliminating the last dimension. The FNN cell in the trunk network takes an input of $N\times2$ coordinates and encodes them into a shape of $101 \times 2$. In addition, the DeepXDE implementation multiplies this output internally $N \times $ in order to predict solutions on all points, making the final encoded output of trunk $N \times 101 \times 2$.  The encoded outputs from both networks are then combined according to \eref{combine} to produce the final outputs. The branch and trunk networks contain 743681 and 62115 trainable parameters, making the total model size 805796. The model was trained for 351000 iterations using an Adam optimizer \cite{kingma2014adam} with a learning rate of $10^{-3}$. Hidden dimensions were set to consist of 100 neurons. The loss function was set to 
modified-$R^2$ as defined in \eref{solidification_loss_func}. 

\begin{equation}
\begin{aligned}
    {\rm{R^2_{Modified}}} = \frac{\sum_{i=1}^{2N} (\Theta_{FE} - \Theta_{Pred})^2}{\frac{n_{Sam}}{2N}\sum_{i=1}^{2N} (\Theta_{FE} - \overline{\Theta_{FE}})^2}
     \label{solidification_loss_func}
\end{aligned}
\end{equation}

In this context, $\Theta_{FE}$ and $\Theta_{Pred}$ represent the outcomes derived from the FEA simulations and predictions made by the neural network, respectively. Furthermore, $N$ is utilized to denote the number of data points (the number of finite element nodes) within a single sample multiplied by the batch size, and $n_{Sam}$ is the batch size, meaning the number of samples in each batch. The reason for extending the summation up to 2N in \eref{solidification_loss_func} stems from the fact that both $\Theta_{FE}$ and $\Theta_{Pred}$ encompass 2N values (stress and temperature at each node). This extension comes from the multiphysics nature of the problem, which incorporates the predictions of both von Mises stress and temperature.

\subsubsection{ResUNet-based DeepONet}
\label{sec:resunet_and_deeponet}
When the geometry of prediction is constantly changing, He et al. \cite{HE2023116277} proposed a variant of the DeepONet architecture with a ResUNet \cite{diakogiannis2020resunet} as the trunk network to capture the variable geometries in a structured grid. With the powerful spatial encoding capabilities of the ResUNet, the authors demonstrated accurate predictions of the von Mises stress field on various topology-optimized geometries under variable loads. In this work, this model was trained using FEA data for LDED AM process generated in \ref{sec:AM_data_generation}, which encompasses huge variety of different geometries. We also extended this architecture to solve multiphysics problem: predict the temperature and residual stress fields of different topology-optimized designs during AM. The architecture was modified accordingly to predict two output components instead of 1 in the original work. This work focuses on the capability of the modified architecture to simultaneously predict two distinct, but entangled properties accurately, and investigates the impact of printing velocity on the prediction outcomes. Therefore the laser power used in AM process was kept as constant for all simulations.

The solution operator $\hat{G}$ is a function of two variables: printing velocity, and topology-optimized domain. It produces a 3-dimensional matrix which has depth of 2, each layer representing a field value of stress and temperature. A comprehensive calculation of the solution operator is in \eref{rdon_operator}. Within this context, bold notations $\boldsymbol{B}$ and $\boldsymbol{T}$ represent output matrices from the branch and trunk network, respectively. The subscript b indicates the batch size, n signifies the flattened geometry dimension, and h denotes the hidden dimension axis over which the summation is performed. The operation concludes with adding a bias $\beta$ to the equation.

\begin{equation}
    {G}_{bnc} = \sum_{h=1}^{HD} {B}_{bch} {T}_{bnh} + {\beta}
    \label{rdon_operator}
\end{equation}

\begin{figure}[h!] 
    \centering
         \includegraphics[width=0.95\textwidth]{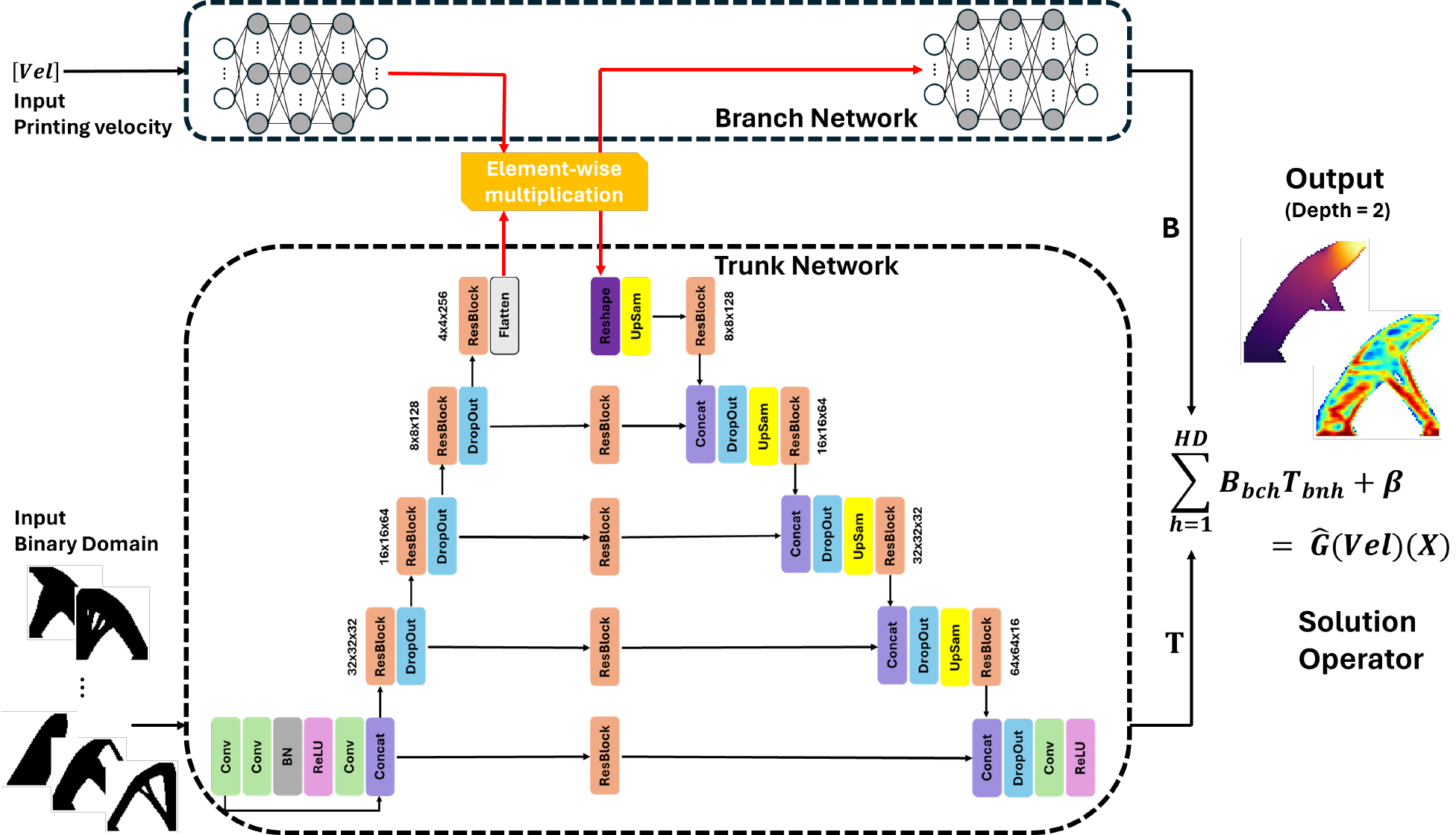}
    \caption{Modified ResUNet-based DeepONet structure for multiphysics problems. \textit{Vel} represent the printing velocity. $\Hat{G}$ represents a solution operator, and $HD$ is the number of hidden dimensions.}
    \label{r_don_structure}
\end{figure}

The network architecture is shown in \fref{r_don_structure}.

The branch network is built to encode the printing velocity, while the ResUNet trunk network encodes the input design images (64, 64). At the end of the ResUNet encoding, when the latent space dimension is the smallest (in this case, 4096), an intermediate data fusion between the branch and trunk networks is introduced via an element-wise product following the work of Wang et al. \cite{wang2022improved}.
The ResUNet-DeepONet architecture used in this work has 2,648,592 trainable parameters. The hidden dimension is set to 32. 4 levels of encoding are adopted in the model, and the initial number of depth channels of the encoding convolution layers was set to 16. The model is trained for 150,000 iterations using the Adam optimizer \cite{kingma2014adam} with an initial learning rate of $5\times 10^{-4}$ to ensure the stability of the learning process convergence. Mean Squared Error (MSE) is utilized as the loss function in this example. The definition of the function is at \eref{AM_loss_func}.

\begin{equation}
\begin{aligned}
    {\rm{MSE}} = \frac{1}{2N}\sum_{i=1}^{2N}(\Theta_{FE} - \Theta_{Pred})^2,\\
     \label{AM_loss_func}
\end{aligned}
\end{equation}

$\Theta_{FE}$ and $\Theta_{Pred}$ denote the results obtained from FEA simulations and the predictions generated by the neural network, respectively. $N$ is employed to represent the quantity of data points, specifically the number of finite element nodes within an individual sample, scaled by the batch size. Along with the previous example in \ref{sec:s-deeponet}, the summation to calculate the loss function goes up to 2N due to the multiphysics nature of the problem; each node contains two physical values, von Mises stress and temperature.

\section{Results and discussion}
\label{sec:results}
All simulations involving physics and finite element analysis were conducted using Abaqus/Standard \citep{Abaqus2022} on an AMD EPYC 7763 Milan CPU core. Additionally, all deep learning training tasks were executed on an Nvidia A100 GPU within Delta, a high-performance computing (HPC) system located at the National Center for Supercomputing Applications (NCSA).

\subsection{Steel Solidification Multi-physics example with S-DeepONet}
\label{sec:s-deeponet_results}

Model performance was evaluated using the conventional 80-20 training-test split. Among 5,494 successful finite element Abaqus simulations, 1,099 cases were randomly selected and used as a test dataset (unseen by the neural network during training). The training process took 5,345 seconds to complete on Nvidia’s A100 GPU. Two test metrics were used in this section to evaluate the NN model performance, as defined in \eref{MAE_def}:

\begin{equation}
\begin{aligned}
    {\text{\text{Mean absolute error}}} = \frac{1}{N} \sum_{i=1}^{N} |\Phi_{FE, i} - \Phi_{Pred, i}| \\
    {\text{\text{CoP}}} = 1-\frac{\sum_{i=1}^{n_{test}}\left \| \Phi_{FE} - \Phi_{Pred} \right \|_2^2}{\sum_{i=1}^{n_{test}}\left \|  \Phi_{FE} - \mu_{\Phi_{FE}}\right \|_2^2}
     \label{MAE_def}
\end{aligned}
\end{equation}

Here, $\Phi$ denotes the physical value obtained either from FEA simulation or NN prediction, presented in its original scale, diverging from the letter $\Theta$ used in earlier loss function formulations. This underscores that these are metric calculations to assess model performance, and they are not utilized as loss functions. Both $\Phi_{FE}$ and $\Phi_{Pred}$ comprise $N$ scalar elements, the number of elements in the finite element simulation. The first metric, mean absolute error (MAE), is evaluated per each test sample. It sums the absolute value of the element-wise differences between the simulation and prediction on a single sample, yielding a total of 1,099 mean absolute errors corresponding to each physical property: temperature and stress. The mean absolute error on each test sample signifies how much the prediction fluctuated from the ground truth in the original scales. 

The Coefficient of Prognosis (CoP) is a model-metric, meaning one scalar value represents the whole model performance \cite{most2011sensitivity}. The maximum CoP can get is up to 1, the case when the prediction perfectly matches the ground truth, the FEA simulation. While $n_{test}$ means the total test samples, the CoP divides the sum of $L_2$ norm between the simulation and prediction of all test cases with respect to total variation. The total variation is calculated as the sum of the $L_2$ norm between the simulation and its mean, also for all test samples. The value of CoP implies how much the prediction varies from the variance of the ground truth. Essentially, CoP tells how close the model's predictions converge toward the mean for values not previously encountered in the dataset.

\begin{figure}[h!] 
    \centering
     \subfloat[Best case]{
         \includegraphics[trim={0cm 0cm 0cm 0cm},clip,width=0.185\textwidth]{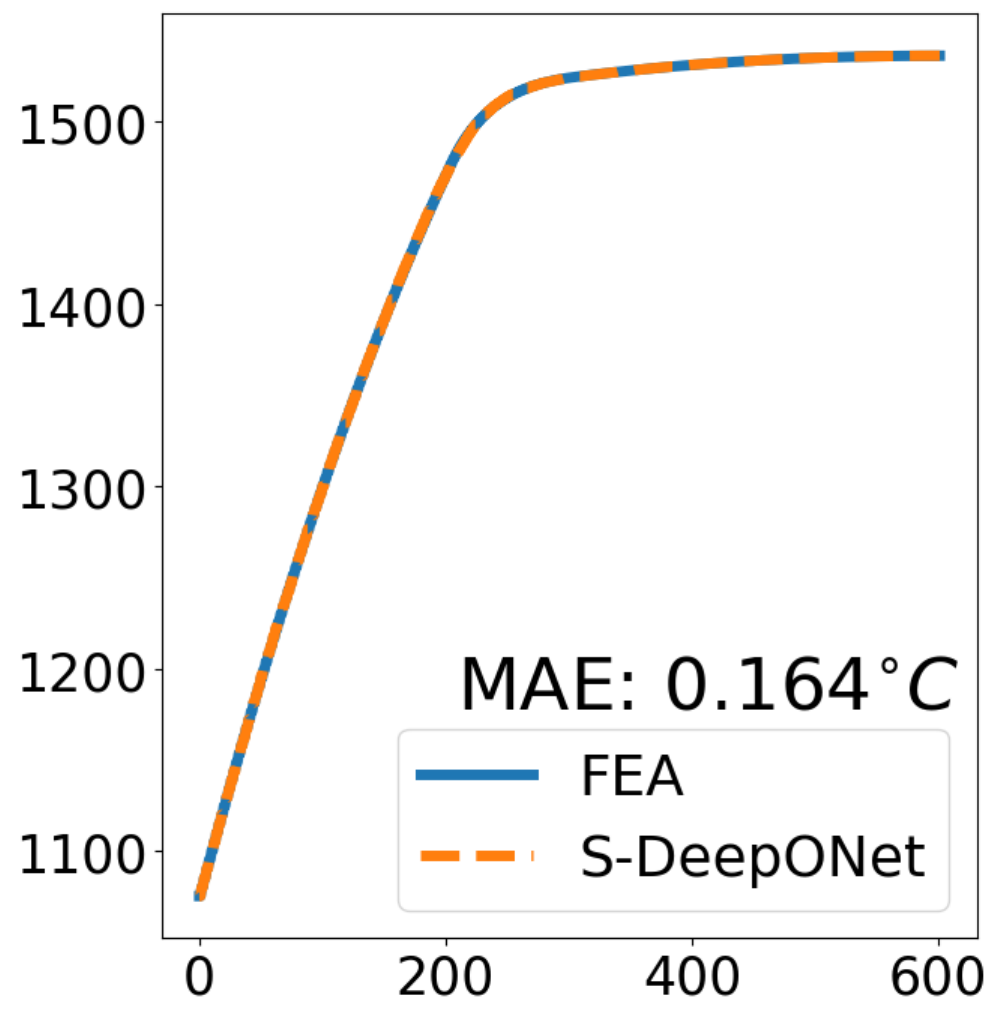}
         \label{temp_0_percentile}
     }
     \subfloat[$25^{th}$ percentile]{
         \includegraphics[trim={0cm 0cm 0cm 0cm},clip,width=0.185\textwidth]{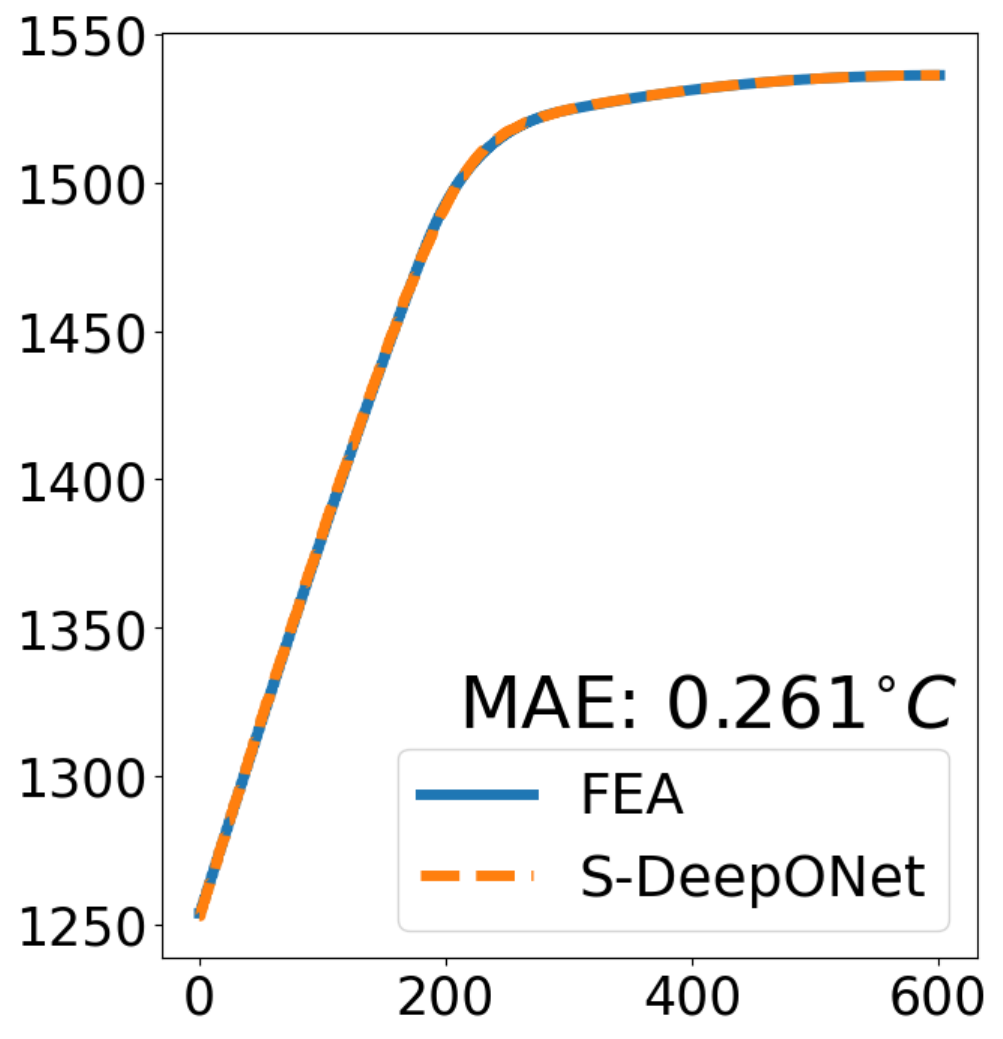}
         \label{temp_25_percentile}
     }
     \subfloat[$50^{th}$ percentile]{
         \includegraphics[trim={0cm 0cm 0cm 0cm},clip,width=0.185\textwidth]{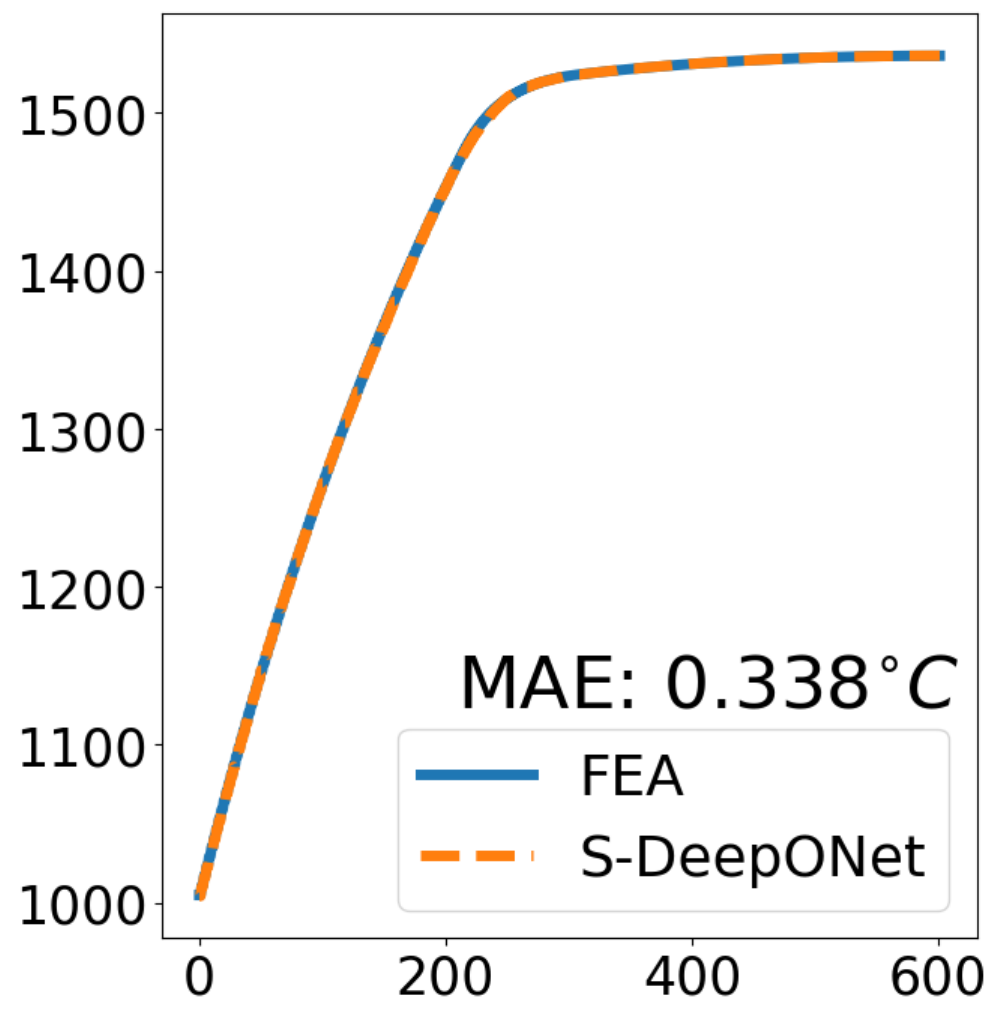}
         \label{temp_50_percentile}
     }
     \subfloat[$75^{th}$ percentile]{
         \includegraphics[trim={0cm 0cm 0cm 0cm},clip,width=0.185\textwidth]{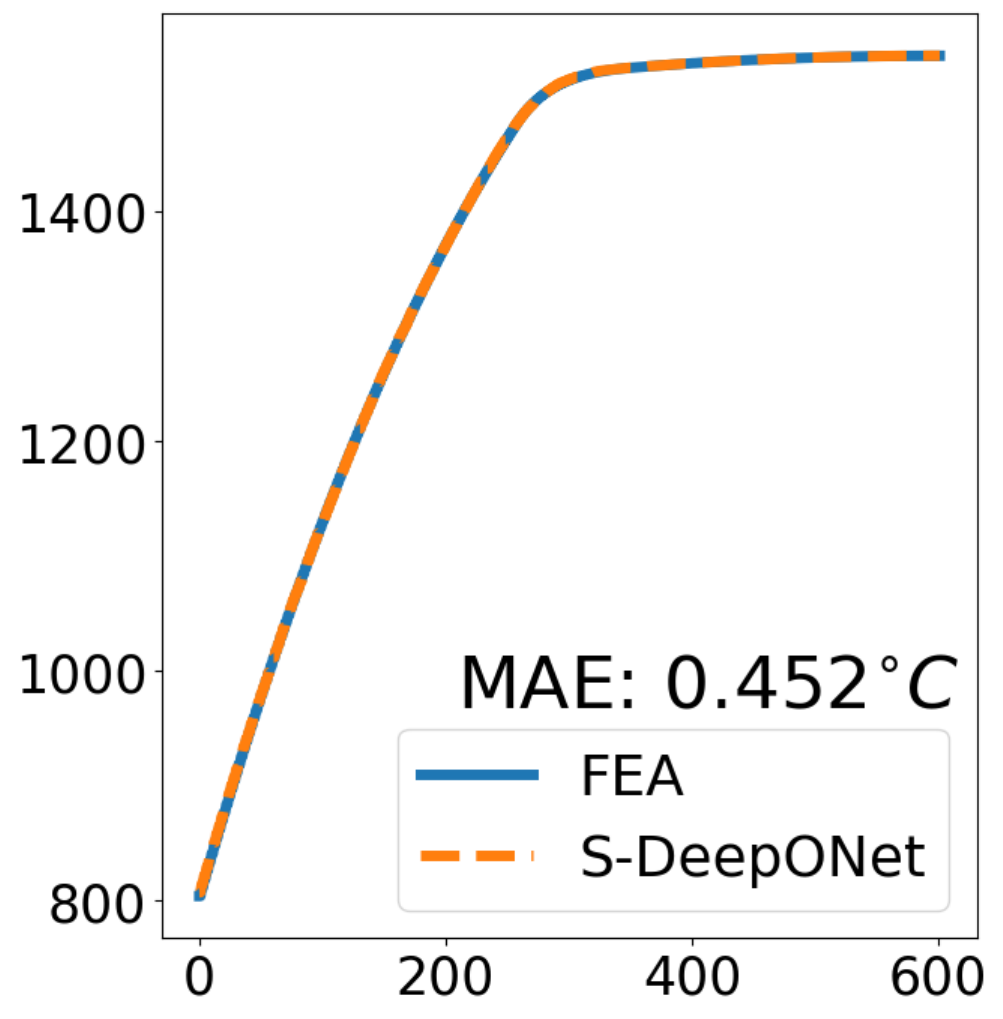}
         \label{temp_75_percentile}
     }
     \subfloat[Worst case]{
         \includegraphics[trim={0cm 0cm 0cm 0cm},clip,width=0.185\textwidth]{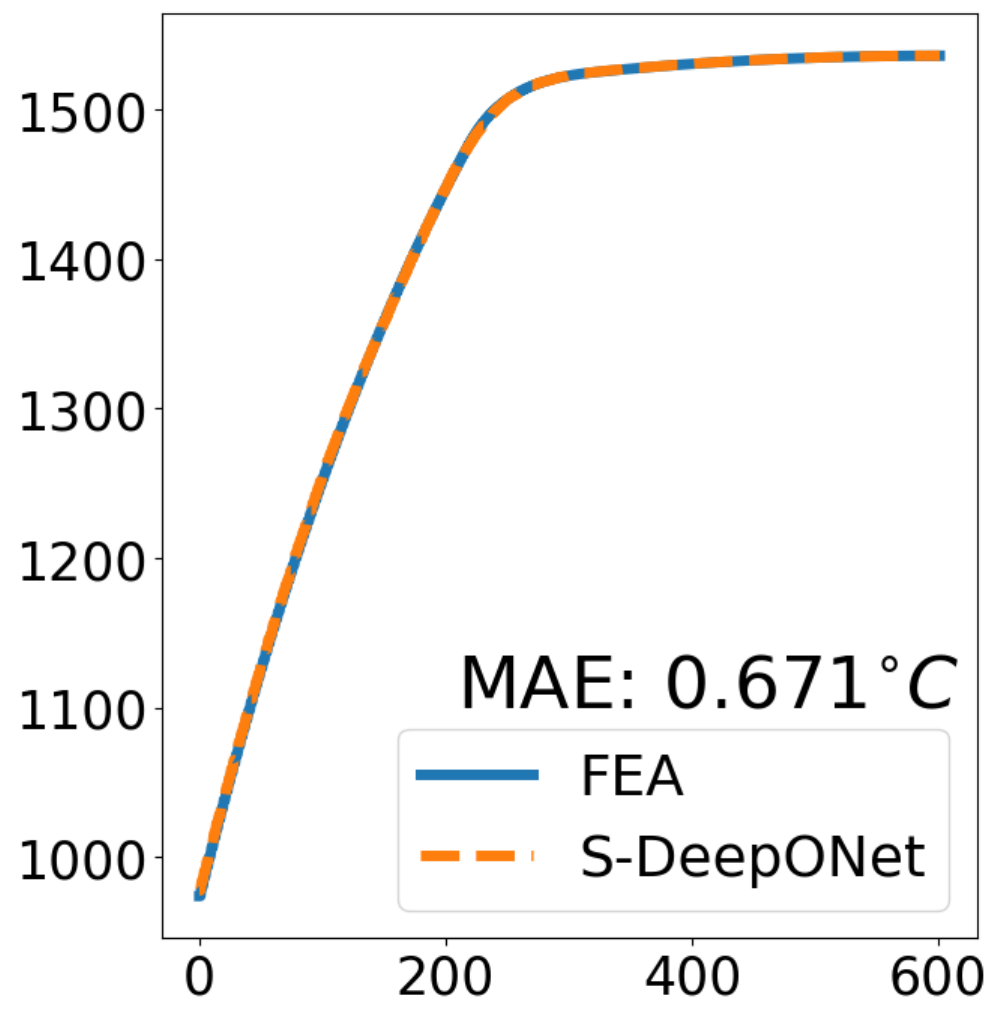}
         \label{temp_90_percentile}
     }
    \caption{Temperature predictions by S-DeepONet compared to multiphysics  FEA simulations in test dataset}
    \label{temp_comp_plots}
\end{figure}

\begin{figure}[h!] 
    \centering
     \subfloat[Best case]{
         \includegraphics[trim={0cm 0cm 0cm 0cm},clip,width=0.185\textwidth]{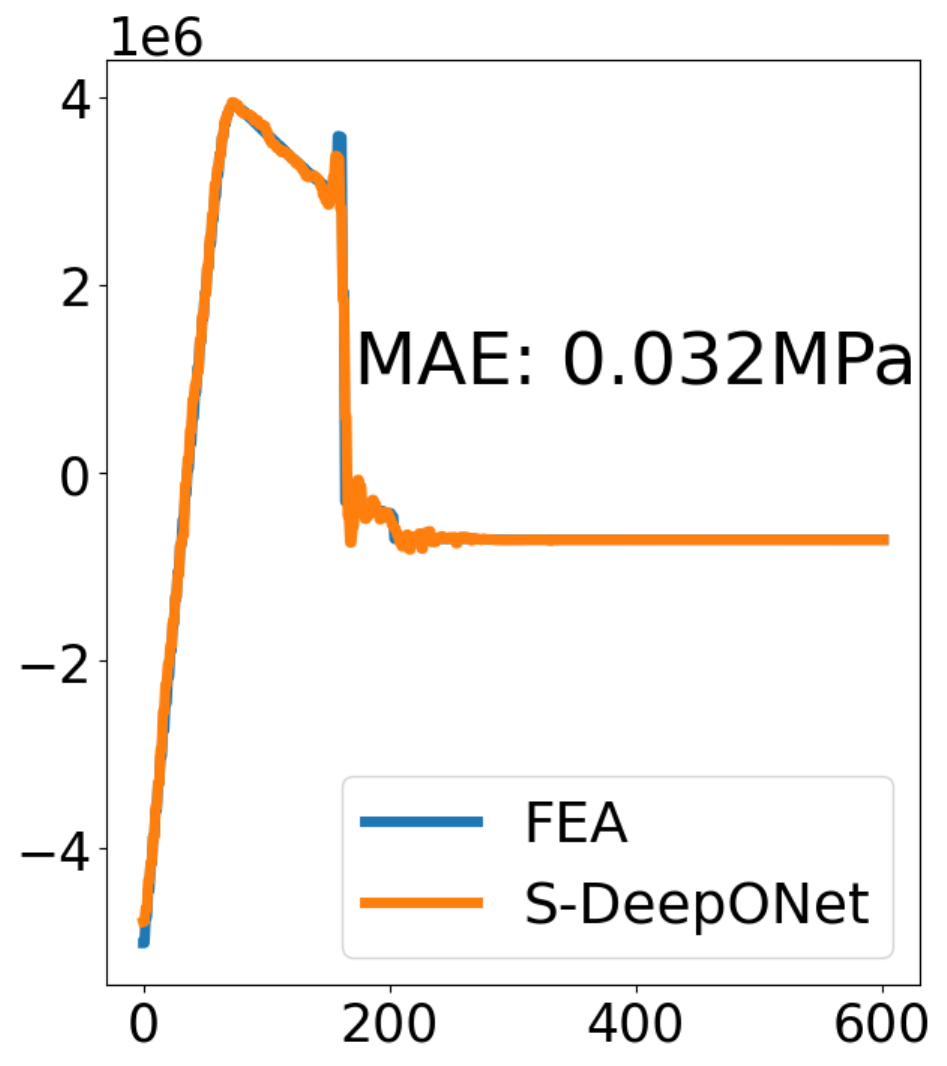}
         \label{stress_0_percentile}
     }
     \subfloat[$25^{th}$ percentile]{
         \includegraphics[trim={0cm 0cm 0cm 0cm},clip,width=0.185\textwidth]{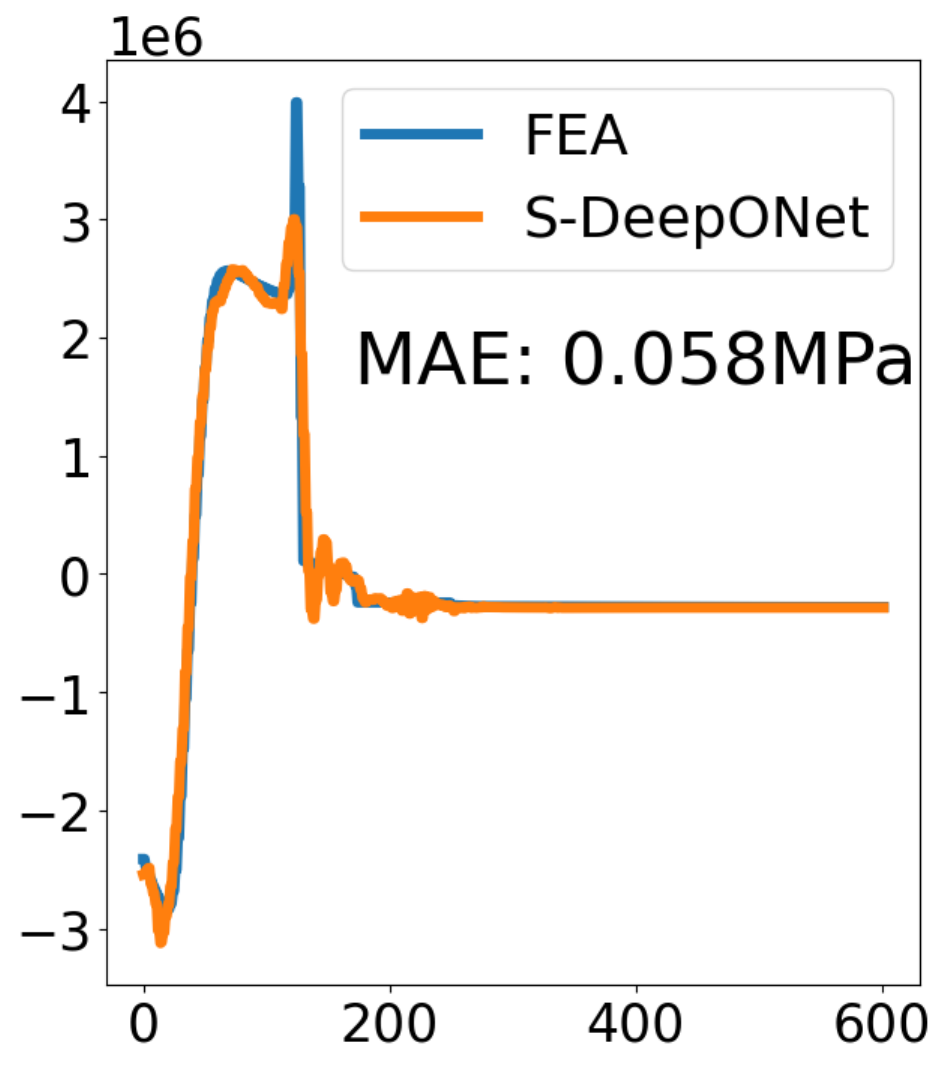}
         \label{stress_25_percentile}
     }
     \subfloat[$50^{th}$ percentile]{
         \includegraphics[trim={0cm 0cm 0cm 0cm},clip,width=0.185\textwidth]{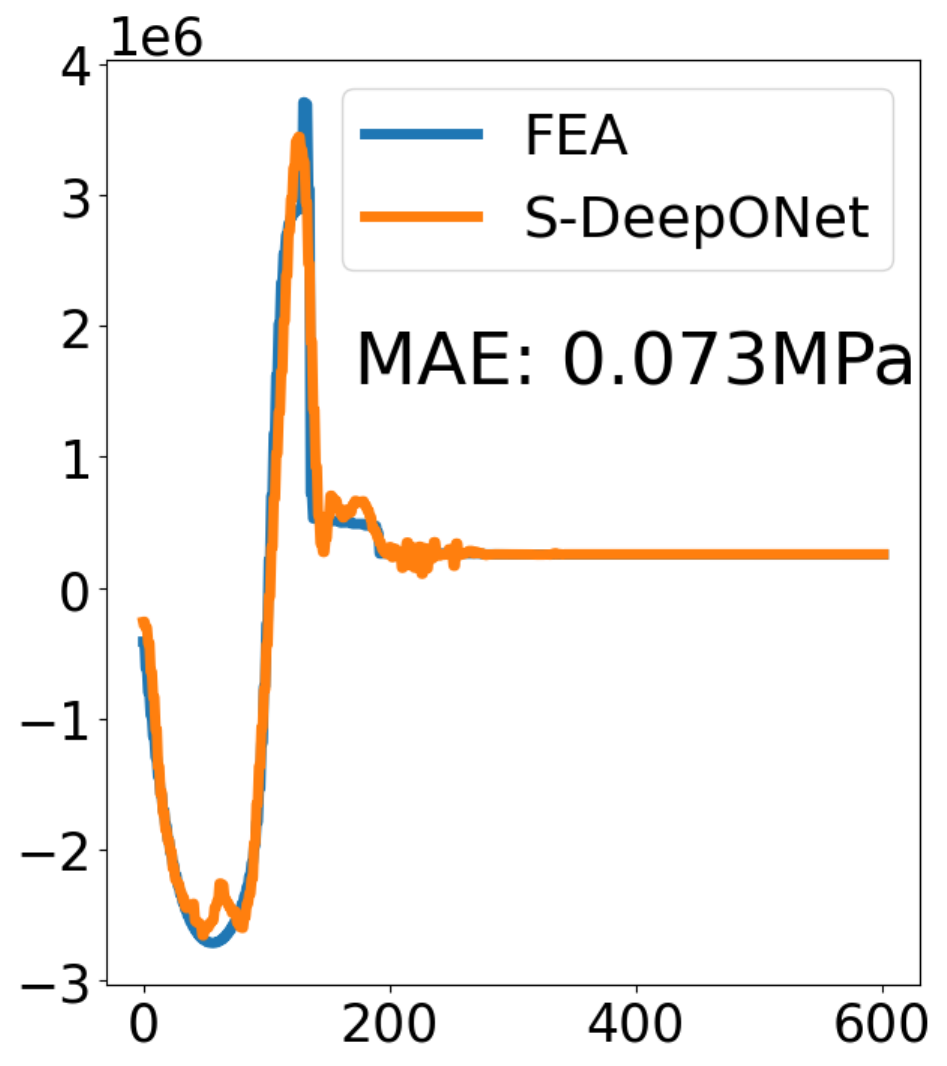}
         \label{stress_50_percentile}
     }
     \subfloat[$75^{th}$ percentile]{
         \includegraphics[trim={0cm 0cm 0cm 0cm},clip,width=0.185\textwidth]{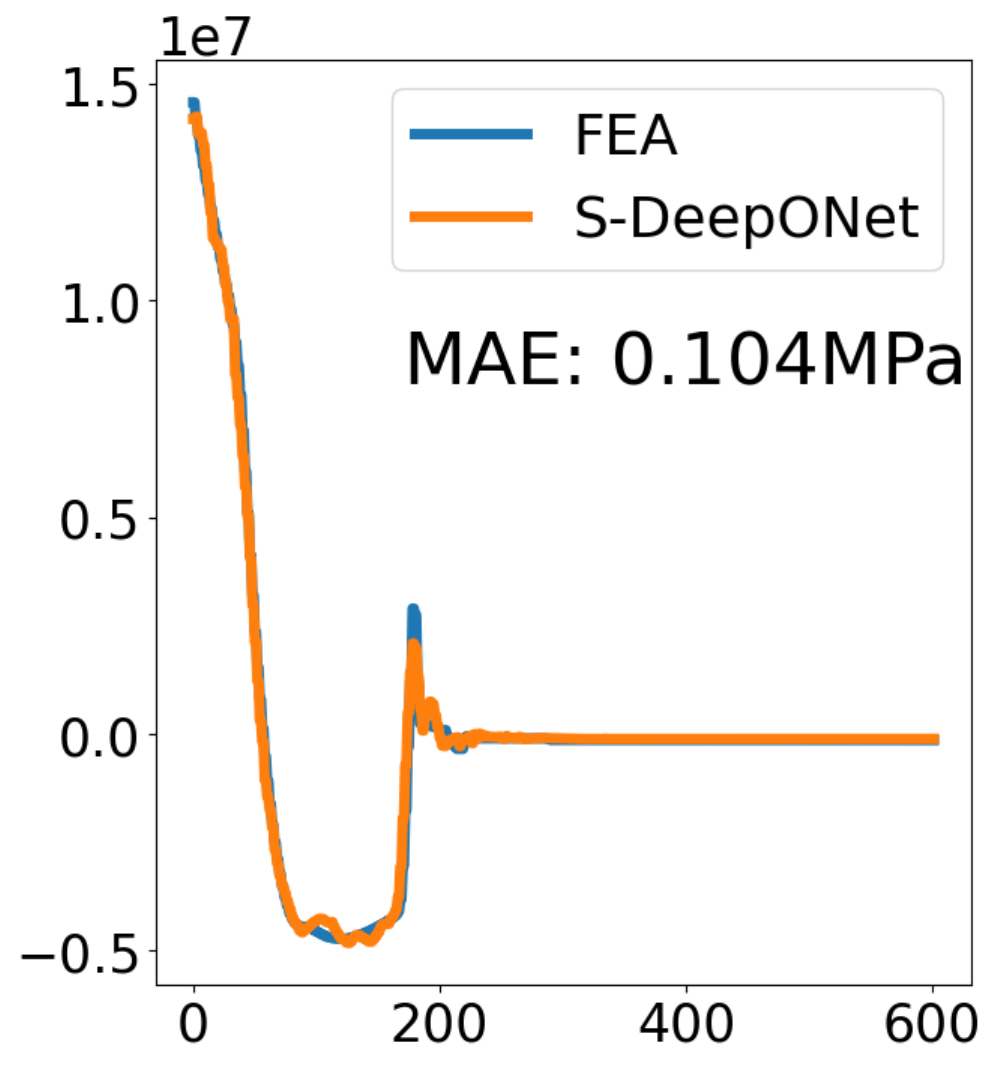}
         \label{stress_75_percentile}
     }
     \subfloat[Worst case]{
         \includegraphics[trim={0cm 0cm 0cm 0cm},clip,width=0.185\textwidth]{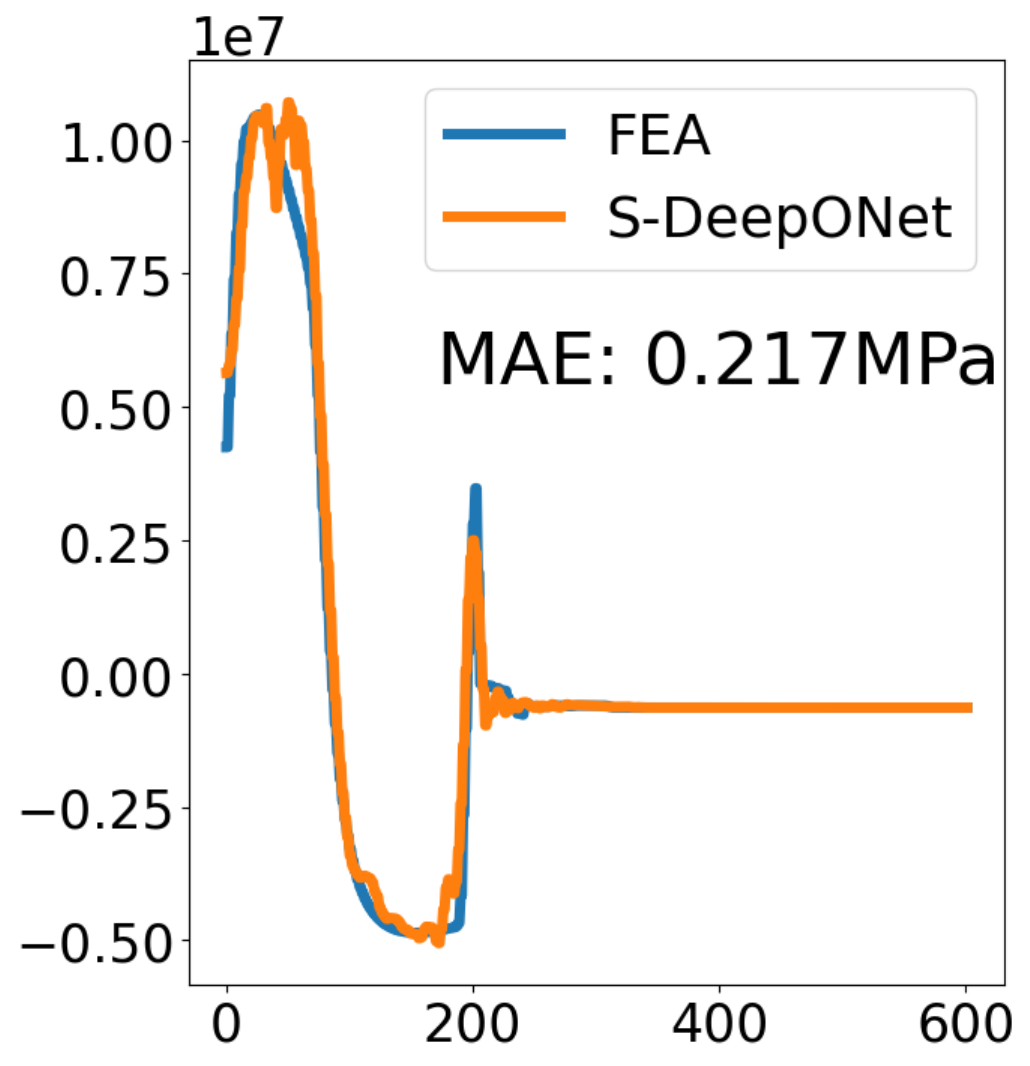}
         \label{stress_90_percentile}
     }
    \caption{Stress predictions by S-DeepONet compared to multiphysics FEA simulations in test dataset}
    \label{stress_comp_plots}
\end{figure}

Due to a large variety of prescribed thermal and displacement histories applied as boundary conditions on the chilled edge, occasionally, solidification is slowed, and the domain remains mostly in the liquid and mushy zones with negligible strength and small, almost constant stress magnitudes. These cases have no relevance in stress-based casting failure predictions but also represent outlier targets that are challenging to predict by any ML methods. For this reason, the absolute error metric provides better interpretation than relative errors, which include division by target norms (small numbers in these cases). Also, to avoid some of these near-zero outliers, the results are presented up to 90\% test error distribution. Temperature distributions, shown in \fref{temp_comp_plots}, are inherently smooth and less challenging to predict than the stress distribution. The worst test case predicted the entire temperature distribution within 0.67$^{\circ}C$ absolute error, which is very accurate considering that the steel temperatures in continuous casting molds are in the 1000$^{\circ}C$ – 1500$^{\circ}C$ range. The CoP value for temperature was 0.998, indicating that the model is stable enough to explain 99.8\% of the data. Comparison of stress predictions and their ground truths are shown in \fref{stress_comp_plots}. Stress distributions are strongly influenced by the 3 constitutive models (austenite, delta-ferrite, and mushy/liquid) based on current phase factions and temperature in the solidification process. In addition, stress depends on the temperature variation through thermal strains and variable displacement history imposed on the chilled surface. This creates a very noisy stress distribution along the slice domain and poses a challenge for accurate prediction.  Despite the obvious difficulties, S-DeepONet managed to predict stress distribution reasonably well, including stress reversals with tensile stress at the chilled surface at 75th \% and the worst test error distributions, which are very important to predict for some failure mechanisms such as hot tearing. The absolute error was in the range of 0.03 – 0.2 MPa, which is again reasonably accurate given that residual stress in typical steel solidification processes in continuous casting ranges from -5MPa (compression) to 5 MPa (tension). The model CoP value for stress was 0.953, meaning that model prediction explains 95.3\% of the total variation. 

Performance metrics for different percentile accuracy levels are shown in \tref{sdeeponet_MAE}. The first row indicates temperature, while the second row describes stress errors.

\begin{figure}[h!] 
    \centering
    \subfloat[Temperature MAE histogram for test data]{
         \includegraphics[trim={0cm 0cm 0cm 0cm},clip,width=0.5\textwidth]{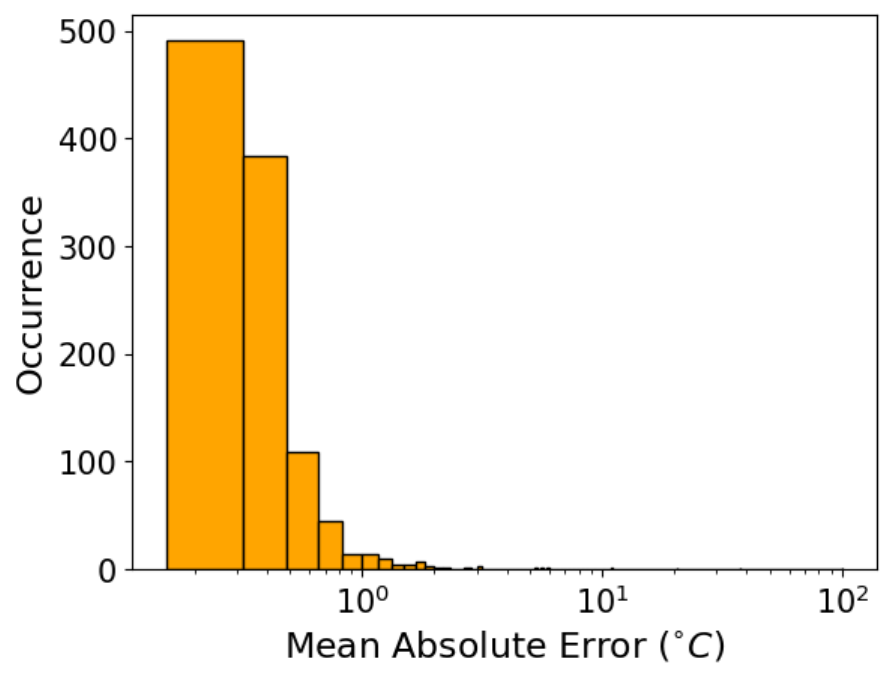}
         \label{temp_hist}
     }
    \subfloat[von Mises stress MAE histogram for test data]{
         \includegraphics[trim={0cm 0cm 0cm 0cm},clip,width=0.5\textwidth]{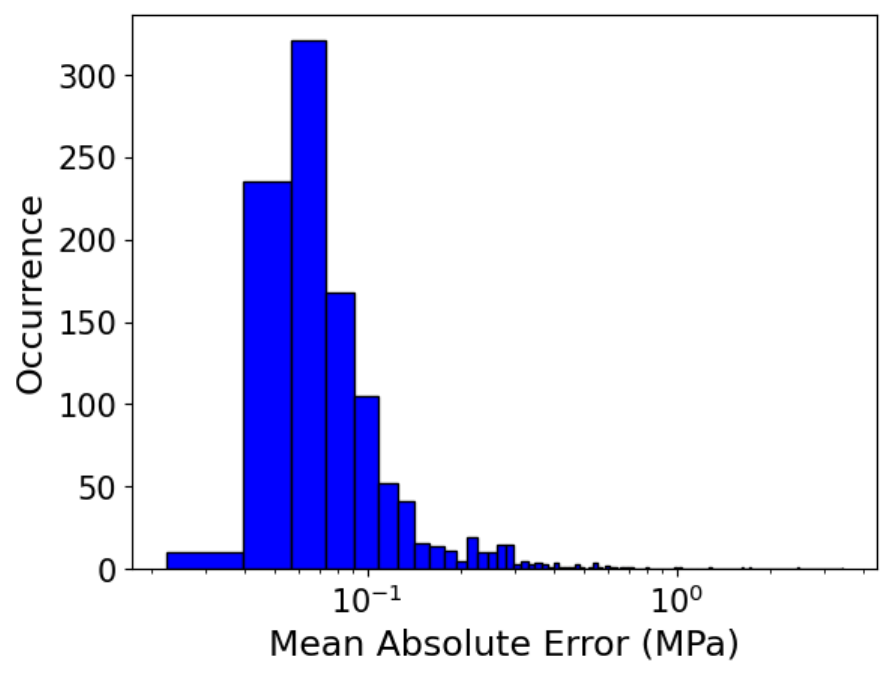}
         \label{stress_hist}
     }
     
    \caption{MAE histograms for temperature and von Mises stress. }
    \label{sdeeponet_hist}
\end{figure}

\fref{sdeeponet_hist} displays a histogram of the absolute errors for temperature and stress test data on a logarithmic scale along the x-axis, illustrating that most test errors cluster closely around the mean (50th \%) absolute errors.

\begin{table}[h!]
    \caption{Detailed prediction statistics of S-DeepONet}
    \small
    \centering
    \begin{tabular}{ccccccccccc}
      & \vline & {\textbf{Best case}} & {\textbf{25$^{th}$ Percentile}} & {\textbf{50$^{th}$ Percentile}} & {\textbf{75$^{th}$ Percentile}} & {\textbf{Worst case}}\\
    \hline
    Temperature MAE ($\bm{^{\circ}C}$) & \vline  & 0.164 & 0.261 & 0.338 & 0.452 & 0.671\\
    Stress MAE (\textbf{MPa}) & \vline  & 0.0320 & 0.058 & 0.073 & 0.104 & 0.217\\
    \end{tabular}
    \label{sdeeponet_MAE}
\end{table}

\begin{figure}[h!] 
    \centering
         \includegraphics[width=0.6\textwidth]{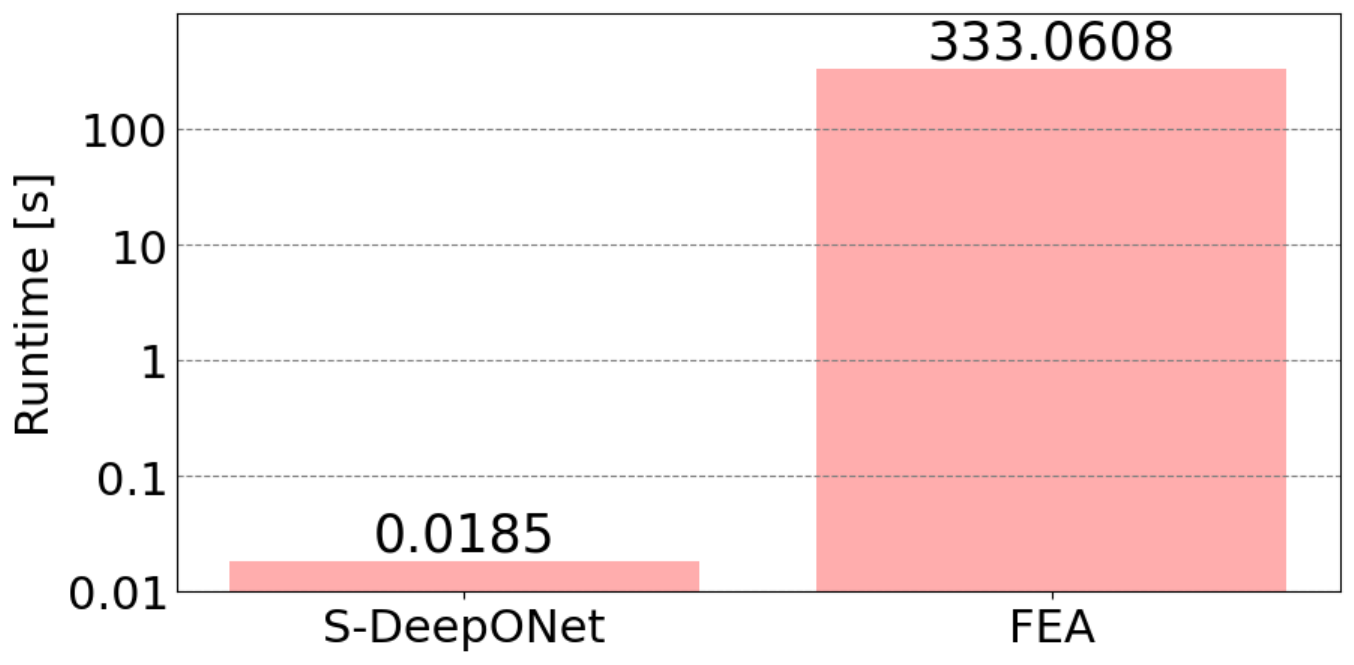}
    \caption{Runtime Comparison Between FEA and S-DeepONet per one sample}
    \label{sdeeponet_speedup}
\end{figure}

The trained S-DeepONet model took 20.3 seconds to predict (infer) all stress and temperature fields for 1,099 test samples, which is 0.0185 seconds to forecast for one unseen sample. Sequentially coupled FEA with highly optimized commercial FEA code and using a parallel solver on the latest HPC took 333 seconds as illustrated in \fref{sdeeponet_speedup}. Conversely, utilizing the developed neural network for prediction demonstrates a remarkable speed improvement, approximately 18,000 times faster.  This efficiency gain highlights the neural network’s capability to provide rapid assessments, significantly reducing the computational time required for classical forward multiphysics evaluations.

\begin{figure}[h!] 
    \centering
     \subfloat[Best case]{
         \includegraphics[trim={0cm 0cm 0cm 0cm},clip,width=0.85\textwidth]{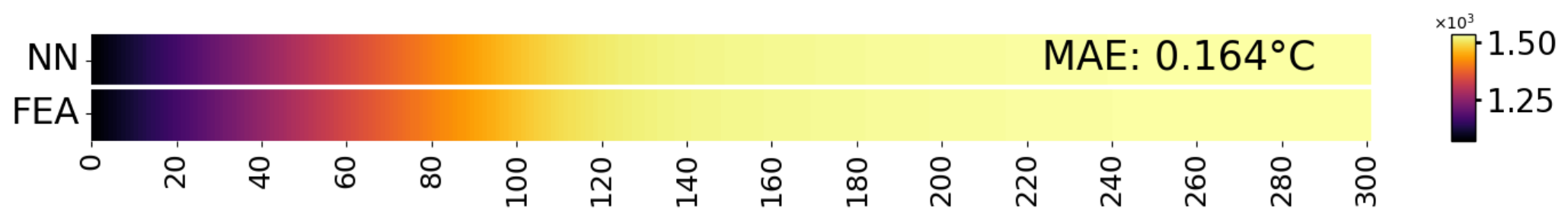}
         \label{temp_contour_0}
     }\\
     \subfloat[$25^{th}$ percentile]{
         \includegraphics[trim={0cm 0cm 0cm 0cm},clip,width=0.85\textwidth]{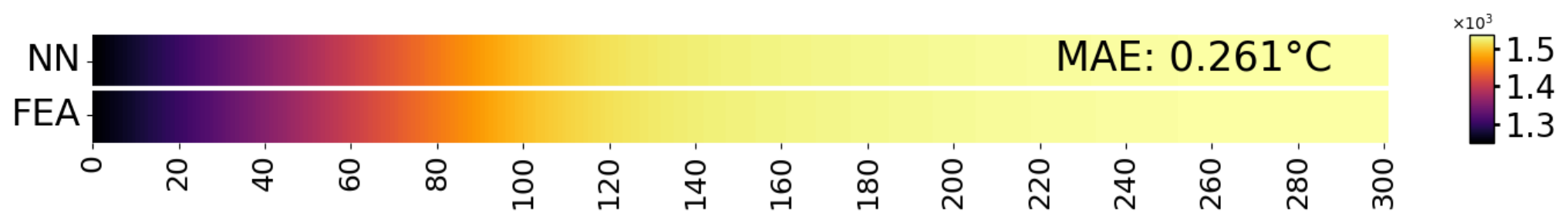}
         \label{temp_contour_25}
     }\\
     \subfloat[$50^{th}$ percentile]{
         \includegraphics[trim={0cm 0cm 0cm 0cm},clip,width=0.85\textwidth]{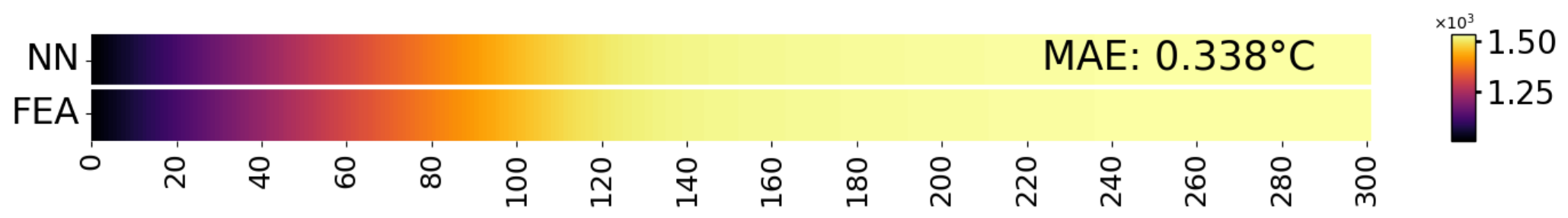}
         \label{temp_contour_50}
     }\\
     \subfloat[$75^{th}$ percentile]{
         \includegraphics[trim={0cm 0cm 0cm 0cm},clip,width=0.85\textwidth]{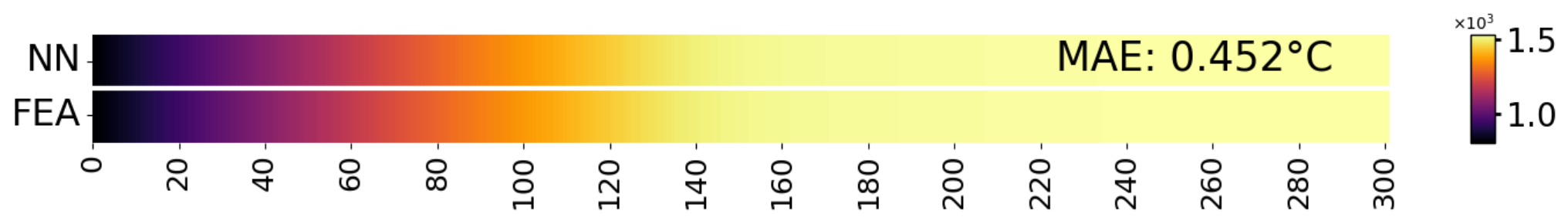}
         \label{temp_contour_75}
     }\\
     \subfloat[Worst case]{
         \includegraphics[trim={0cm 0cm 0cm 0cm},clip,width=0.85\textwidth]{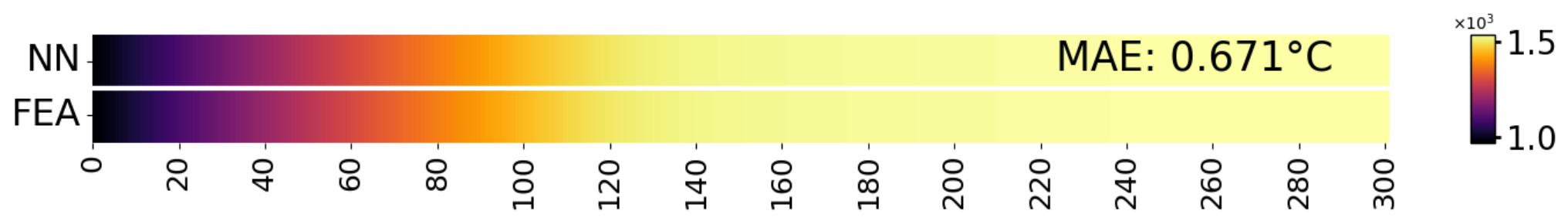}
         \label{temp_contour_90}
     }
    \caption{S-DeepONet temperature contour predictions compared to FEA in the test dataset. NN denotes the S-DeepONet prediction, and FEA denotes the finite element simulation.}
    \label{temp_contours}
\end{figure}

\begin{figure}[h!] 
    \centering
     \subfloat[Best case]{
         \includegraphics[trim={0cm 0cm 0cm 0cm},clip,width=0.85\textwidth]{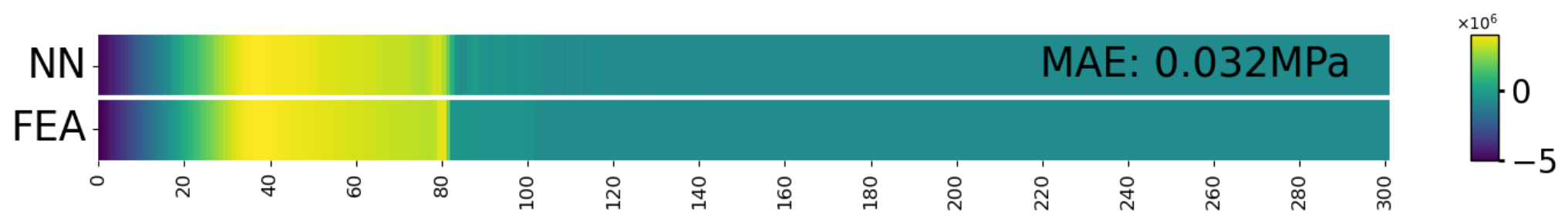}
         \label{stress_contour_0}
     }\\
     \subfloat[$25^{th}$ percentile]{
         \includegraphics[trim={0cm 0cm 0cm 0cm},clip,width=0.85\textwidth]{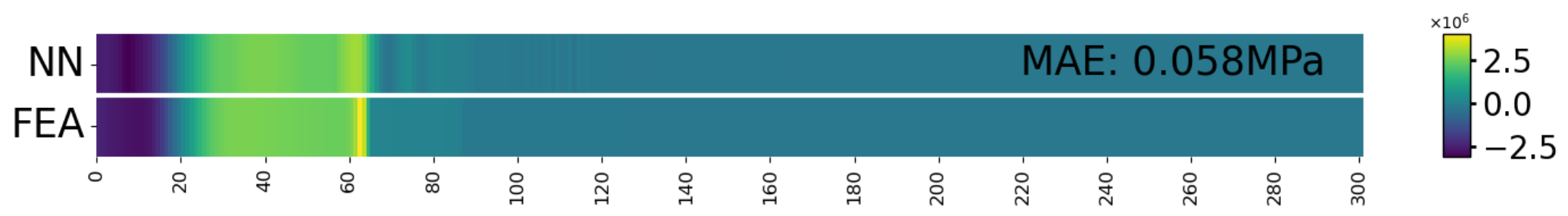}
         \label{stress_contour_25}
     }\\
     \subfloat[$50^{th}$ percentile]{
         \includegraphics[trim={0cm 0cm 0cm 0cm},clip,width=0.85\textwidth]{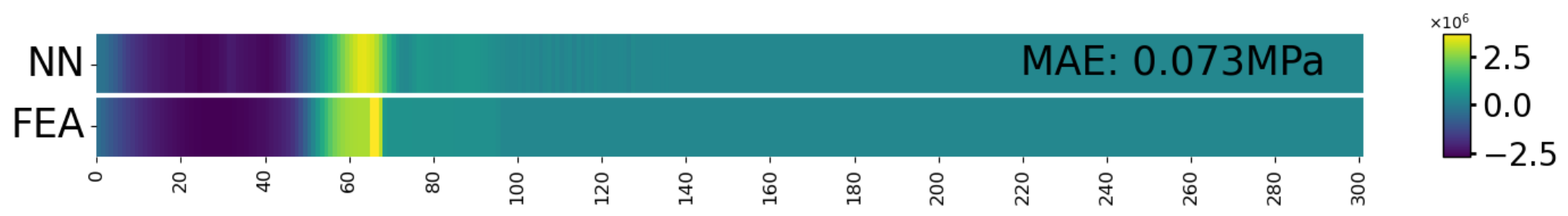}
         \label{stress_contour_50}
     }\\
     \subfloat[$75^{th}$ percentile]{
         \includegraphics[trim={0cm 0cm 0cm 0cm},clip,width=0.85\textwidth]{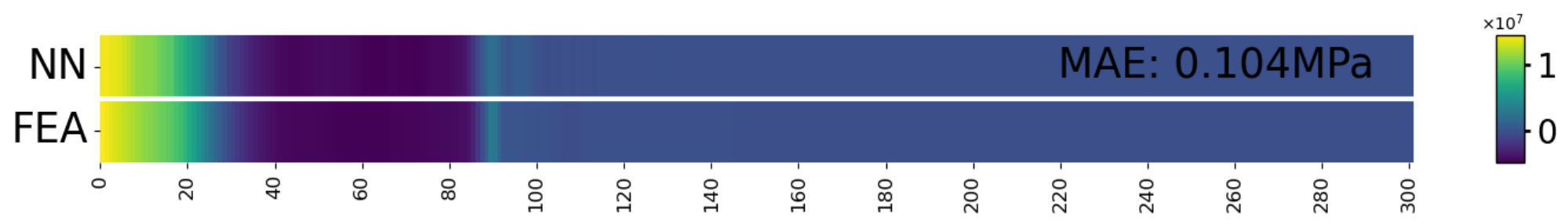}
         \label{stress_contour_75}
     }\\
     \subfloat[Worst case]{
         \includegraphics[trim={0cm 0cm 0cm 0cm},clip,width=0.85\textwidth]{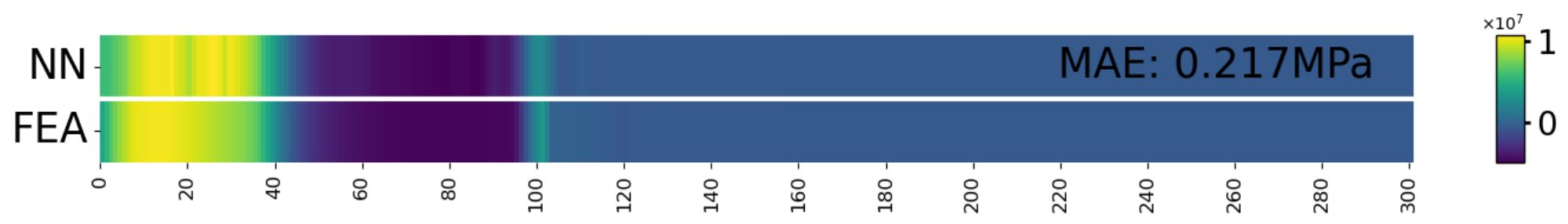}
         \label{stress_contour_90}
     }
    \caption{S-DeepONet stress contour predictions compared to FEA in the test dataset. NN denotes the S-DeepONet prediction, and FEA denotes the finite element simulation.}
    \label{stress_contours}
\end{figure}

Finally, \fref{temp_contours} and \fref{stress_contours} display the physical representations for temperature and stress, respectively. The comparative visual analysis between the predictions made by S-DeepONet and the simulation results from FEA is depicted through these heat maps. In each cases, the top row—labeled NN (Neural Network)—displays the predicted values, whereas the bottom row—labeled FEA—presents the results from finite element simulations, serving as the benchmark for accuracy.

\clearpage

\subsection{Additive Manufacturing Multi-physics example with ResUNet-based DeepONet}
\label{sec:r_don_results}
Model performance was tested using the 80-20 training-test split. Out of 4,574 completed finite element Abaqus simulations, a subset of 915 samples was randomly chosen to be the test dataset. The training process, composed of 150,000 iterations, took 1440 seconds on Nvidia's A100 GPU, making each iteration take 0.0096 seconds. For the evaluation of the trained neural network, 5 metrics in \eref{model_performance_metric} are used: Relative L2 Error, Mean Absolute Error (MAE), Mean Relative $L_2$ Error (MRL2E), Mean Relative Absolute Error (MRAE), and the Coefficient of Prognosis (CoP) \cite{most2011sensitivity}.

\begin{equation}
\begin{aligned}
    \text{Relative $L_2$ error (\%)} = \frac{\left\| \Phi_{\text{FE}} - \Phi_{\text{Pred}} \right\|_2}{\left\| \Phi_{\text{FE}} \right\|_2} \times 100\%\\
    {\text{\text{Mean absolute Error}}} = \frac{1}{N_{points}} \sum_{i=1}^{N_{points}} |\Phi_{FE, i} - \Phi_{Pred, i}| \\
    {\text{MRL2E}(\%)} = \frac{1}{n_{test}}\sum_{i=1}^{n_{test}}\frac{\left \| \Phi _{FE} - \Phi _{Pred} \right \|_2}{\left \| \Phi _{FE} \right \|_2}\times 100 (\%)\\
    {\text{MRAE}(\%)} = \frac{1}{n_{test}}\sum_{i=1}^{n_{test}}\frac{\sum_{j=1}^{N_{points}}\left | \Phi_{FE} - \Phi_{Pred} \right |}{\sum_{j=1}^{N_{points}}\left | \Phi_{FE} - \mu_{\Phi_{FE}} \right |}\times 100 (\%)\\
    {\text{CoP}} = 1-\frac{\sum_{i=1}^{n_{test}}\left \| \Phi_{FE} - \Phi_{Pred} \right \|_2^2}{\sum_{i=1}^{n_{test}}\left \|  \Phi_{FE} - \mu_{\Phi_{FE}}\right \|_2^2}
     \label{model_performance_metric}
\end{aligned}
\end{equation}

Within the context of evaluating the test dataset, which consists of $n_{test}$ samples, the first two metrics outlined in \eref{model_performance_metric} Relative \(L_2\) Error and MAE are computed for each test sample, yielding $n_{test}$ distinct values for each metric per one property. These metrics serve the purpose of assessing performance on an individual samples, facilitating the identification of samples where the model exhibits the highest and lowest performance. $N_{Points}$ represents the number of elements in one sample in both ground truths and predictions. On the other hand, the last three metrics in \eref{model_performance_metric} MRL2E, MRAE, and CoP are a model-metric, which results a scalar value for each model per one physical property. Thus, for one given trained DeepONet model, each metric yields two distinct values, one for von Mises stress and another for temperature. These values serve to quantify the model's predictive accuracy for each physical property.

The results presented hereafter are from the test samples, after manually excluding some extreme designs that cannot be printed out from the powder bed fusion manufacturing process. Therefore, the worst cases in the figures below are 100$^{th}$ percentile data after excluding such designs that are unable to be manufactured.

\begin{table}[h!]
    \caption{Error statistics for temperature}
    \small
    \centering
    \begin{tabular}{ccccccc}
      & \vline 
      & {\textbf{Best case}} 
      & {\textbf{25$^{th}$ Percentile}}
      & {\textbf{50$^{th}$ Percentile}}
      & {\textbf{75$^{th}$ Percentile}}
      & {\textbf{Worst case}} \\
    \hline
    Rel. $L_2$ error (\textbf{\%}) 
    & \vline  
    & 1.25 
    & 2.13 
    & 2.81 
    & 3.72 
    & 8.31\\
    
    MAE (\bm{$^{\circ}C$}) 
    & \vline  
    & 3.62 
    & 5.47 
    & 7.67 
    & 9.71
    & 10.3\\
    
    \end{tabular}
    \label{rdon_temp_samples}
\end{table}

\begin{figure}[h!] 
    \centering
     \subfloat[Best case]{
         \includegraphics[trim={0cm 0cm 0cm 0cm},clip,width=0.20\textwidth]{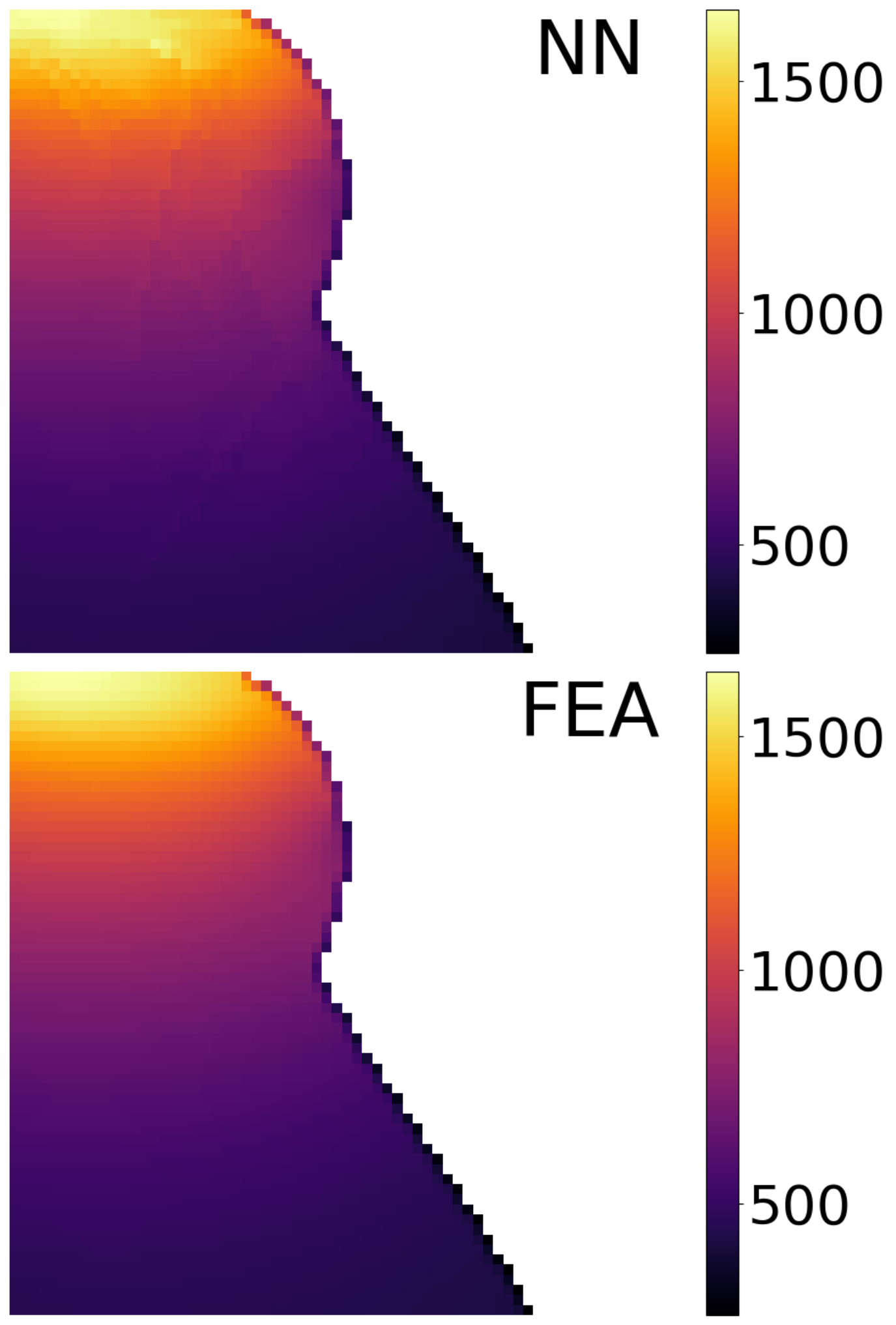}
         \label{rdon_temp_contour_0}
     }
     \subfloat[$25^{th}$ percentile]{
         \includegraphics[trim={0cm 0cm 0cm 0cm},clip,width=0.20\textwidth]{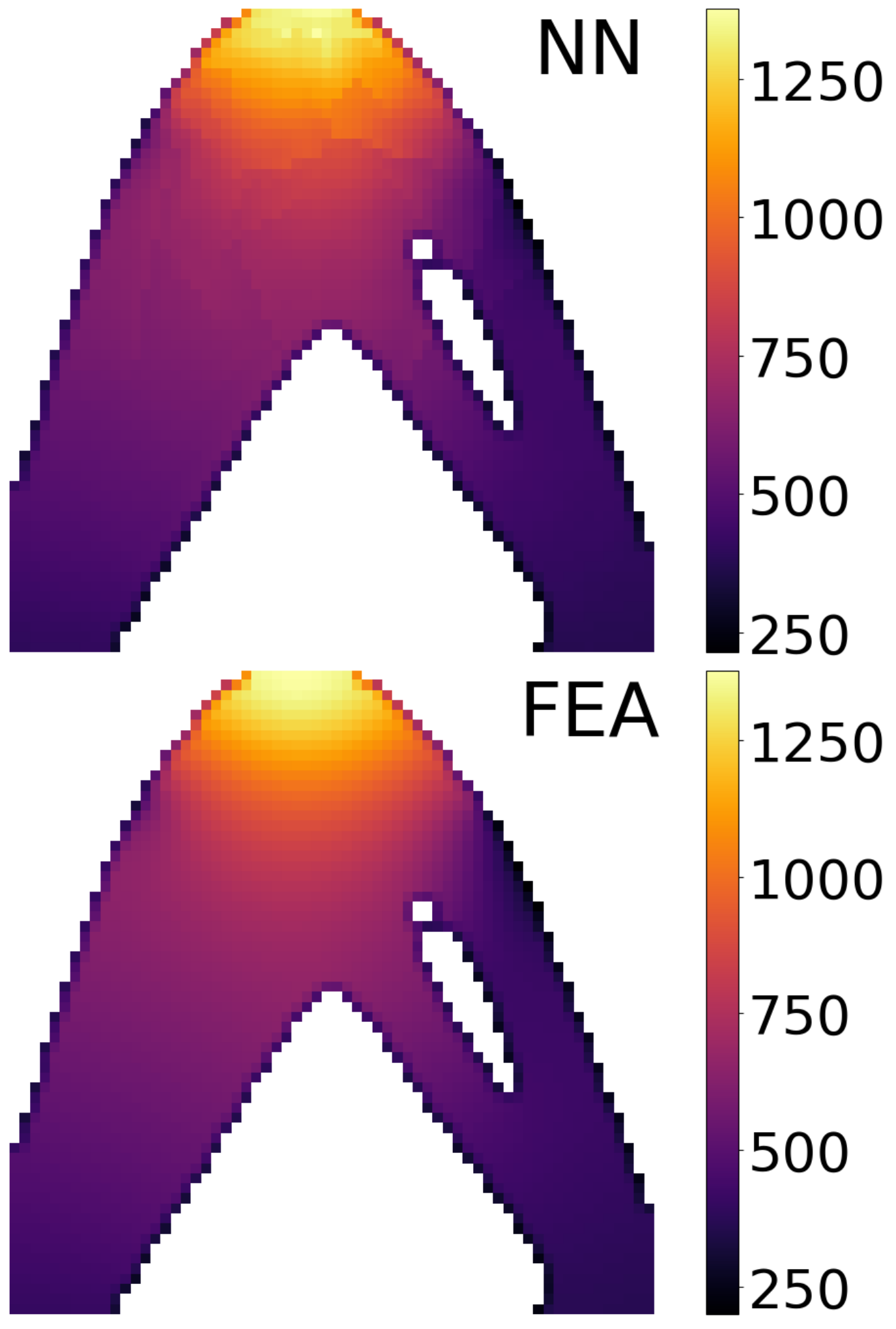}
         \label{rdon_temp_contour_25}
     }
     \subfloat[$50^{th}$ percentile]{
         \includegraphics[trim={0cm 0cm 0cm 0cm},clip,width=0.20\textwidth]{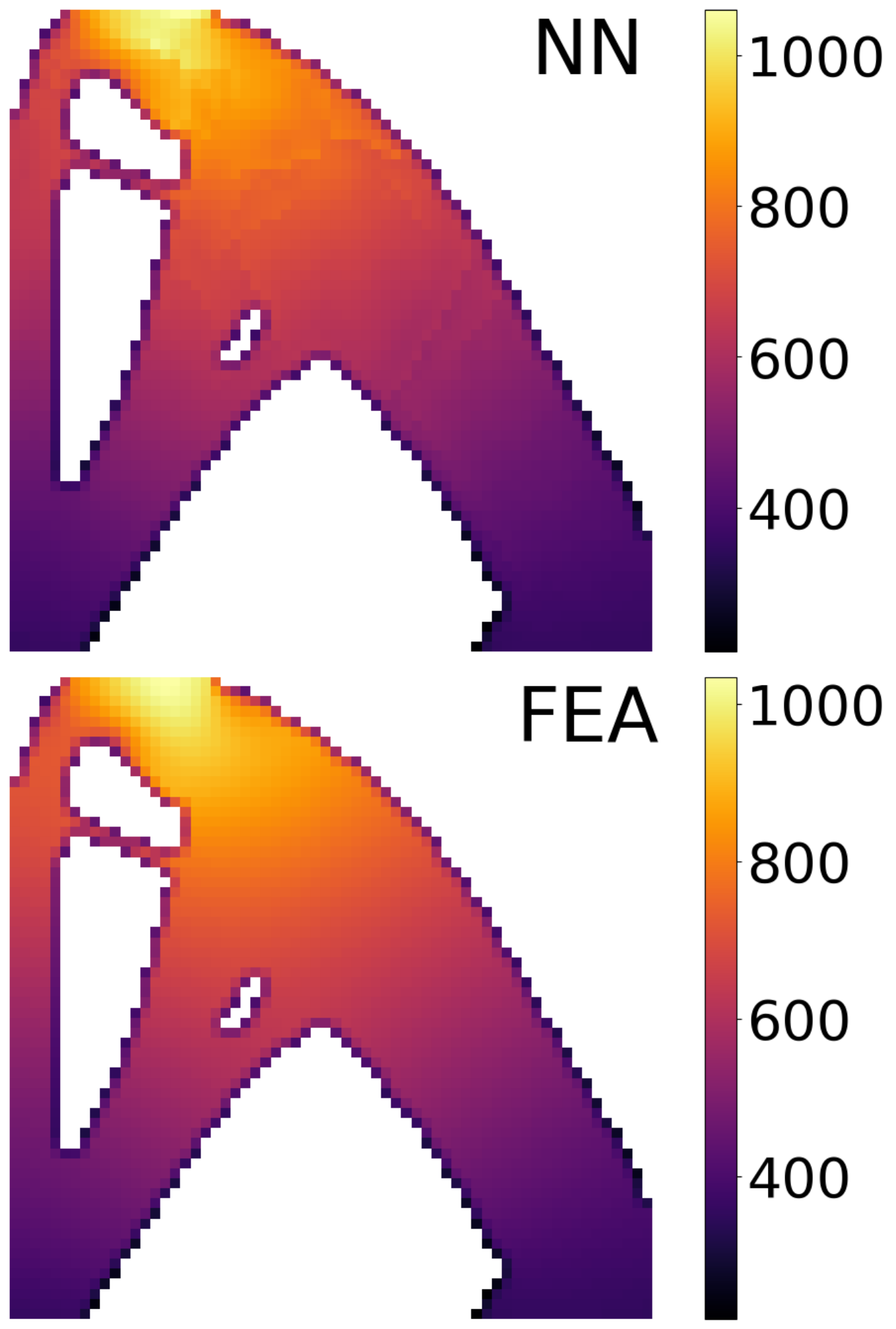}
         \label{rdon_temp_contour_50}
     }
     \subfloat[$75^{th}$ percentile]{
         \includegraphics[trim={0cm 0cm 0cm 0cm},clip,width=0.20\textwidth]{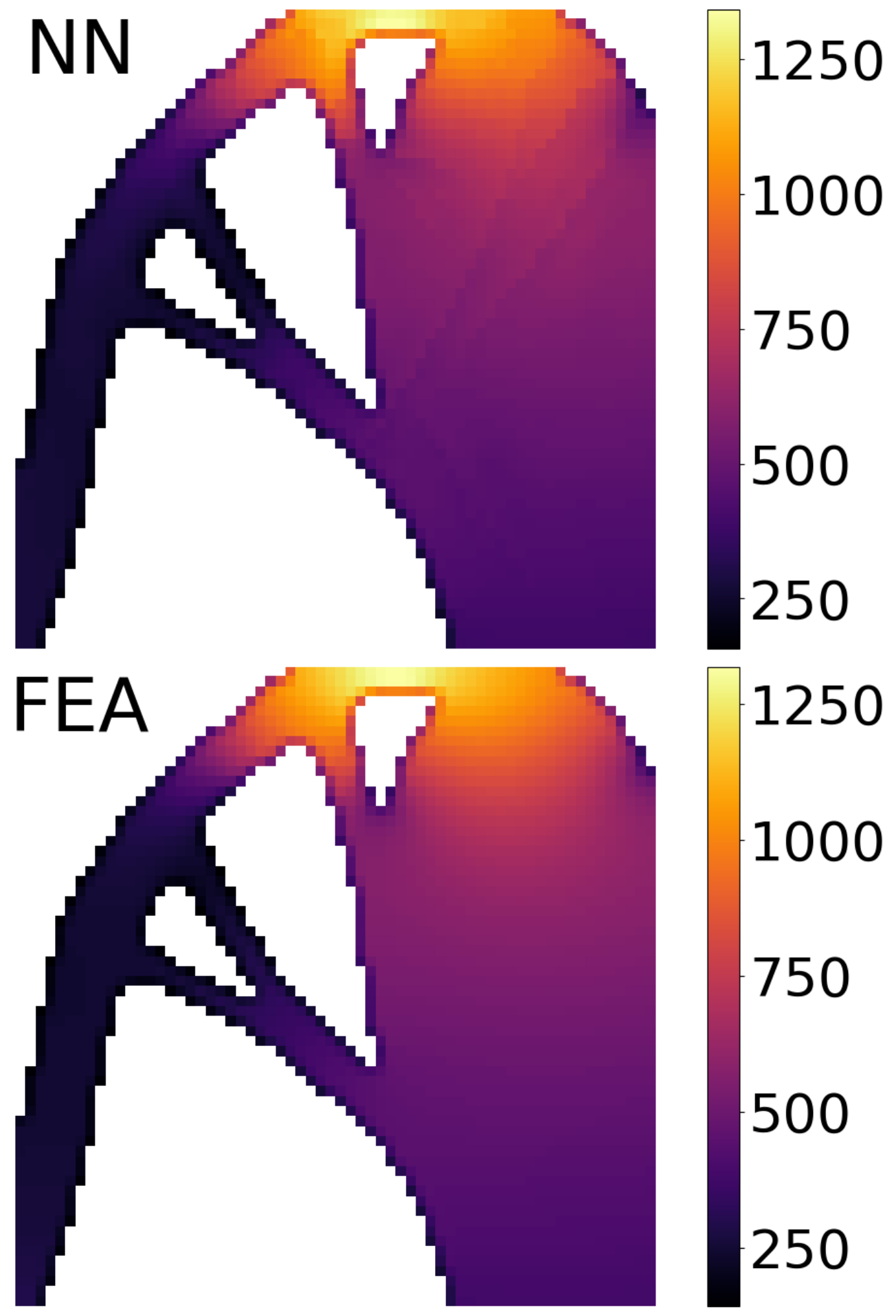}
         \label{rdon_temp_contour_75}
     }
     \subfloat[Worst case]{
         \includegraphics[trim={0cm 0cm 0cm 0cm},clip,width=0.20\textwidth]{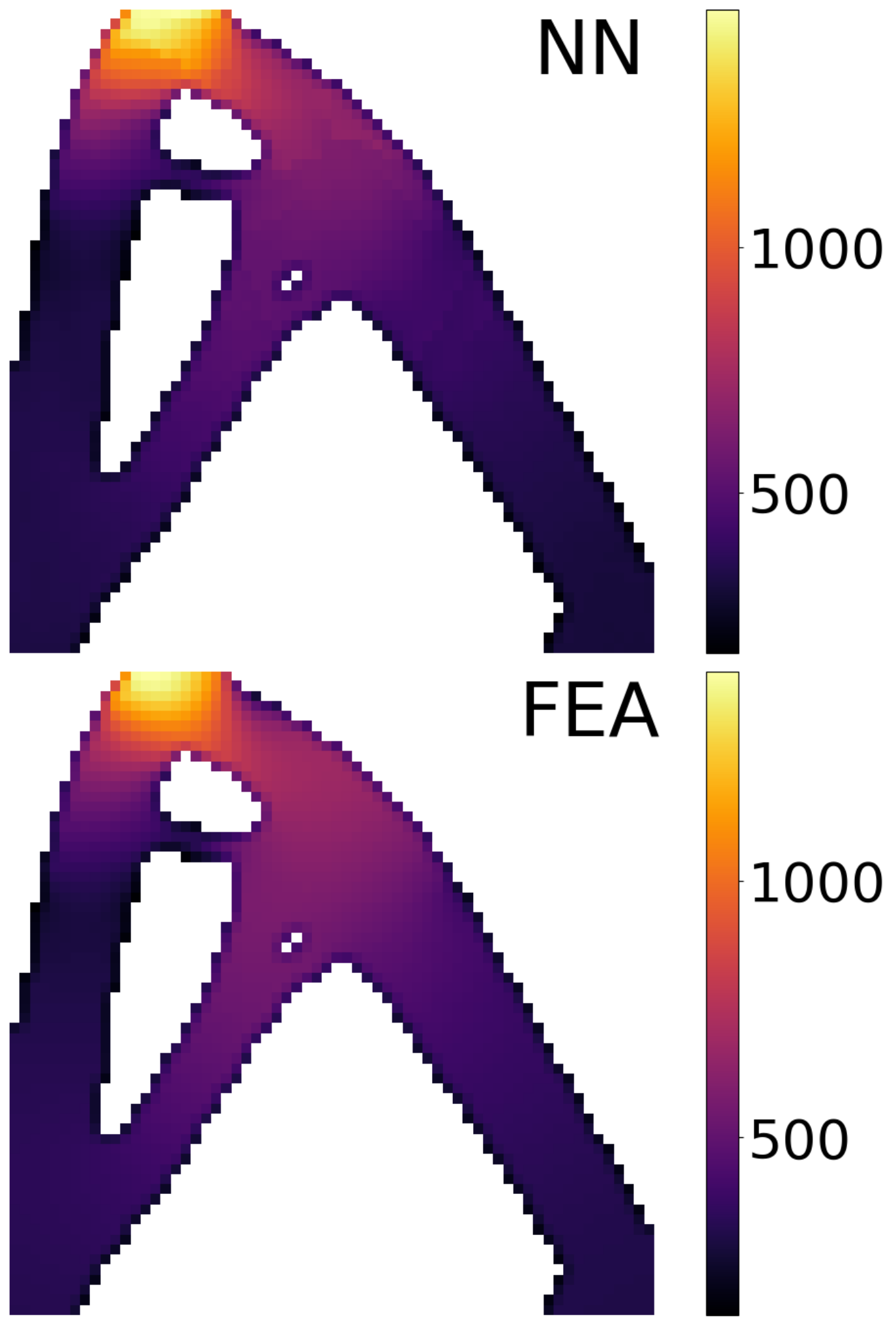}
         \label{rdon_temp_contour_100}
     }
    \caption{Temperature predictions by ResUNet-based DeepONet compared to FEA for samples in \tref{rdon_temp_samples}. (Rel. $L_2$ Error)}
    \label{rdon_temp_contour}
\end{figure}

\tref{rdon_temp_samples} summarizes the ResUNet-based DeepONet performance on the temperature prediction. Relative $L_2$ error was calculated for test samples, which the model has not seen in the training process. Due to complexities in geometry, temperatures became more challenging to predict compared to the previous example in \ref{sec:s-deeponet_results}. Still, the worst case displayed the relative $L_2$ error of 8.16\% as illustrated in \fref{rdon_temp_contour}. "NN" represents the predicted stress field outcomes from ResUNet-based DeepONet model, whereas "FEA" is used to denote the outcomes acquired from finite element analysis.

Upon all test samples, the Mean Relative L2 Error (MRL2E) was recorded at 3.14\%, the Mean Relative Absolute Error (MRAE) at 2.31\%, the average MAE at 6.37$^{\circ}$C, and the CoP reached 0.998 in temperature predictions. The performance metrics of the temperature prediction model are detailed in \tref{rdon_results}.

\begin{figure}[h!] 
    \centering
     \subfloat[Best case]{
         \includegraphics[trim={0cm 0cm 0cm 0cm},clip,width=0.20\textwidth]{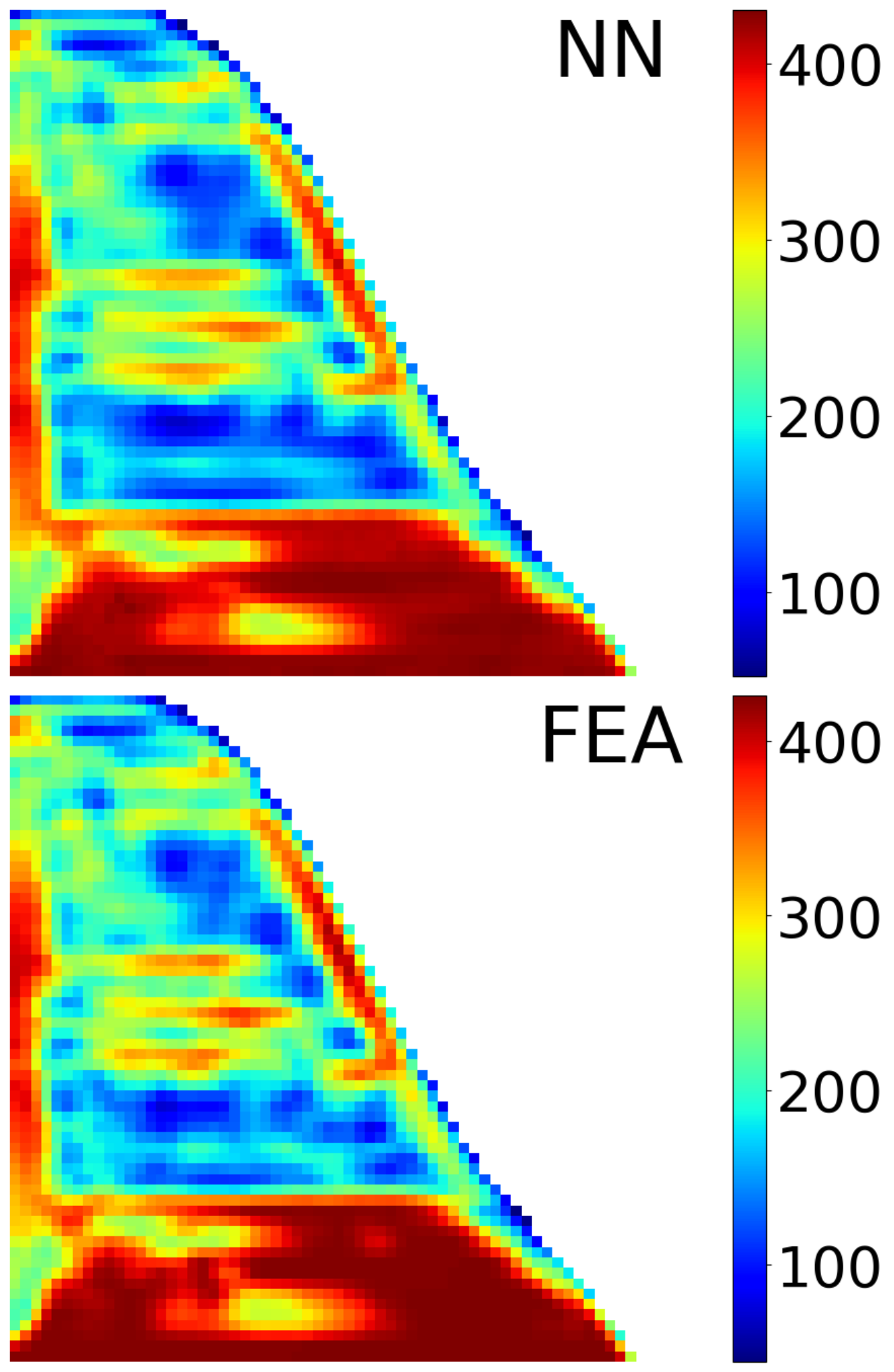}
         \label{rdon_stress_contour_0}
     }
     \subfloat[$25^{th}$ percentile]{
         \includegraphics[trim={0cm 0cm 0cm 0cm},clip,width=0.20\textwidth]{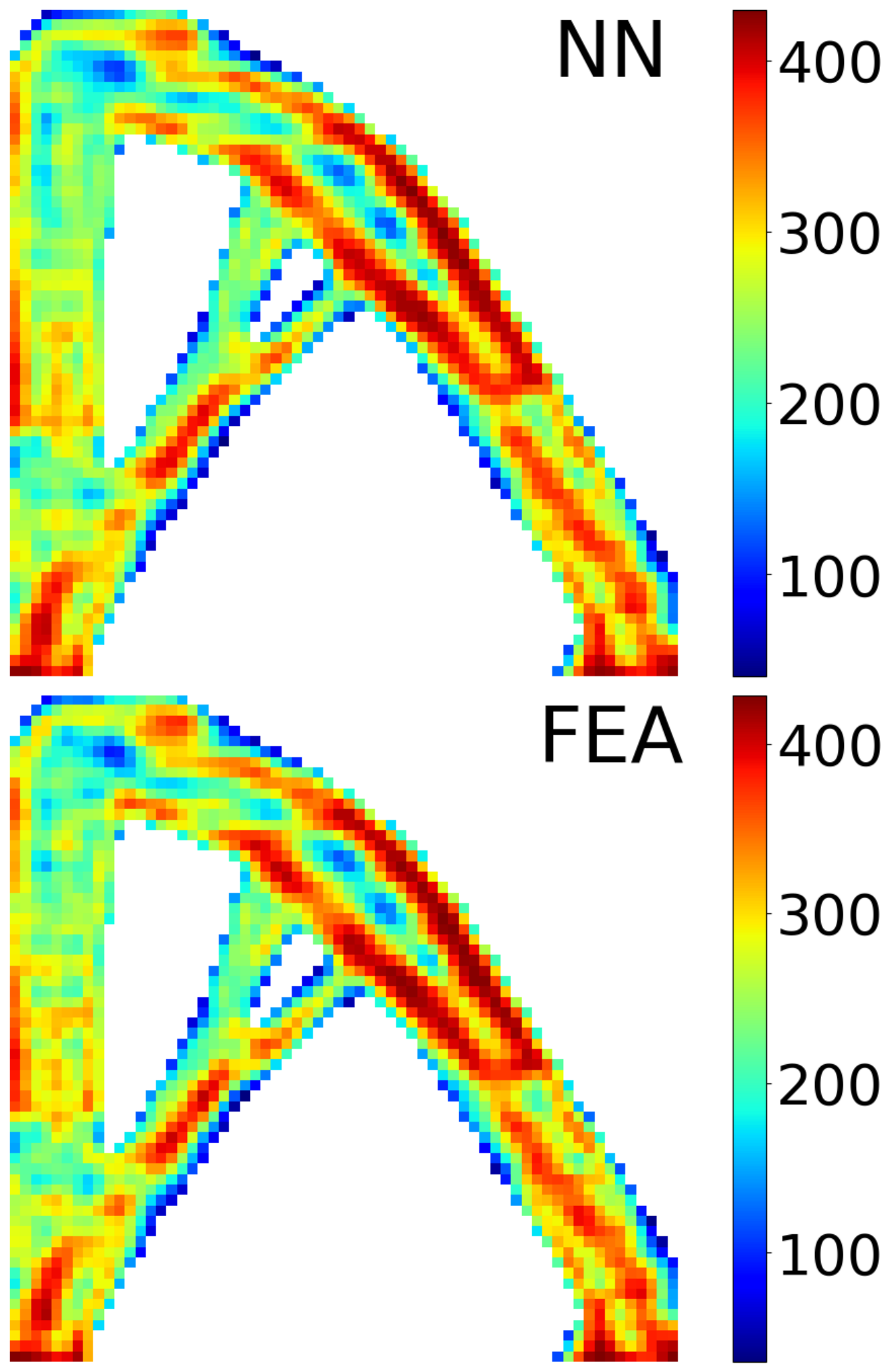}
         \label{rdon_stress_contour_25}
     }
     \subfloat[$50^{th}$ percentile]{
         \includegraphics[trim={0cm 0cm 0cm 0cm},clip,width=0.20\textwidth]{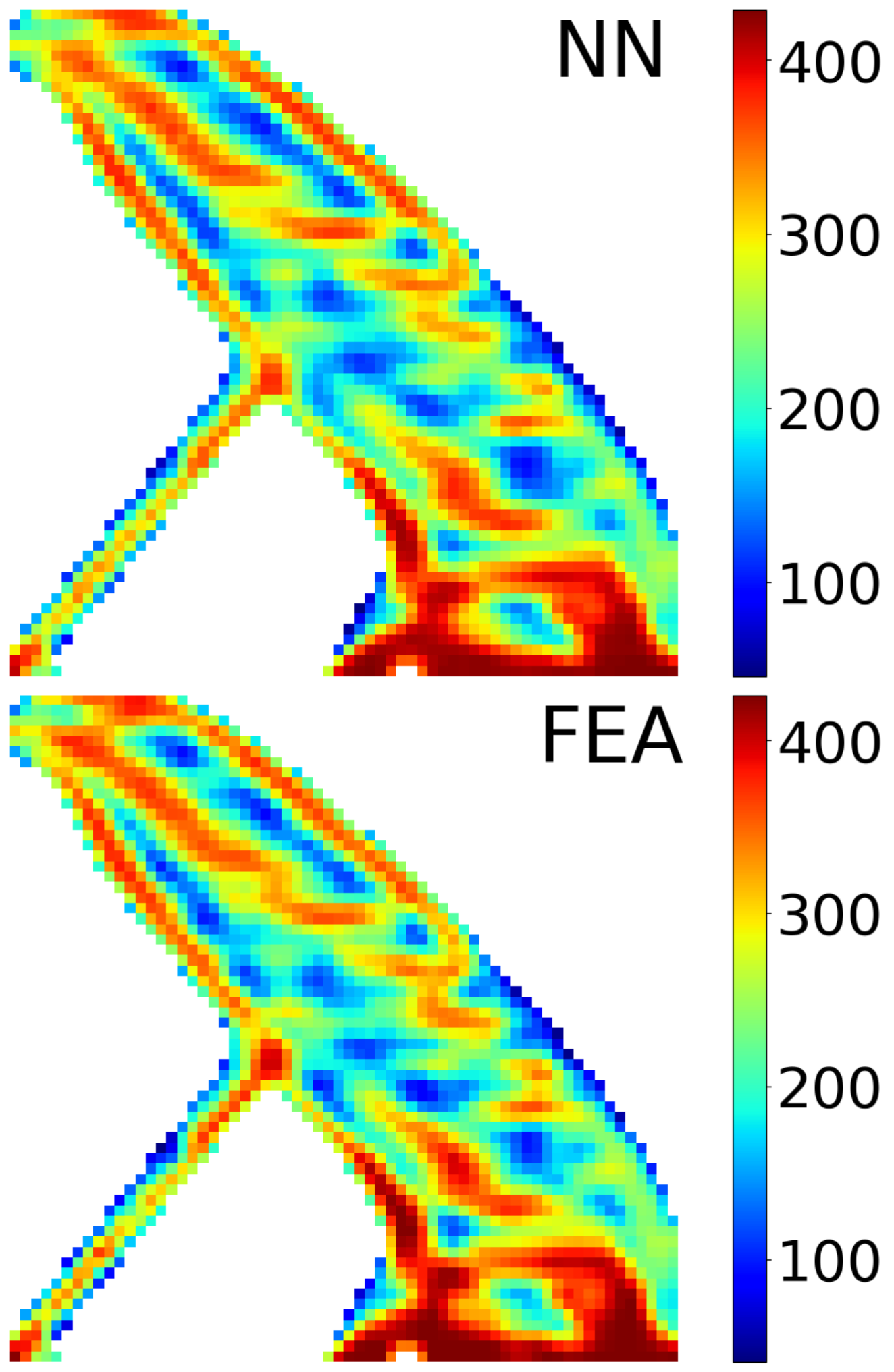}
         \label{rdon_stress_contour_50}
     }
     \subfloat[$75^{th}$ percentile]{
         \includegraphics[trim={0cm 0cm 0cm 0cm},clip,width=0.20\textwidth]{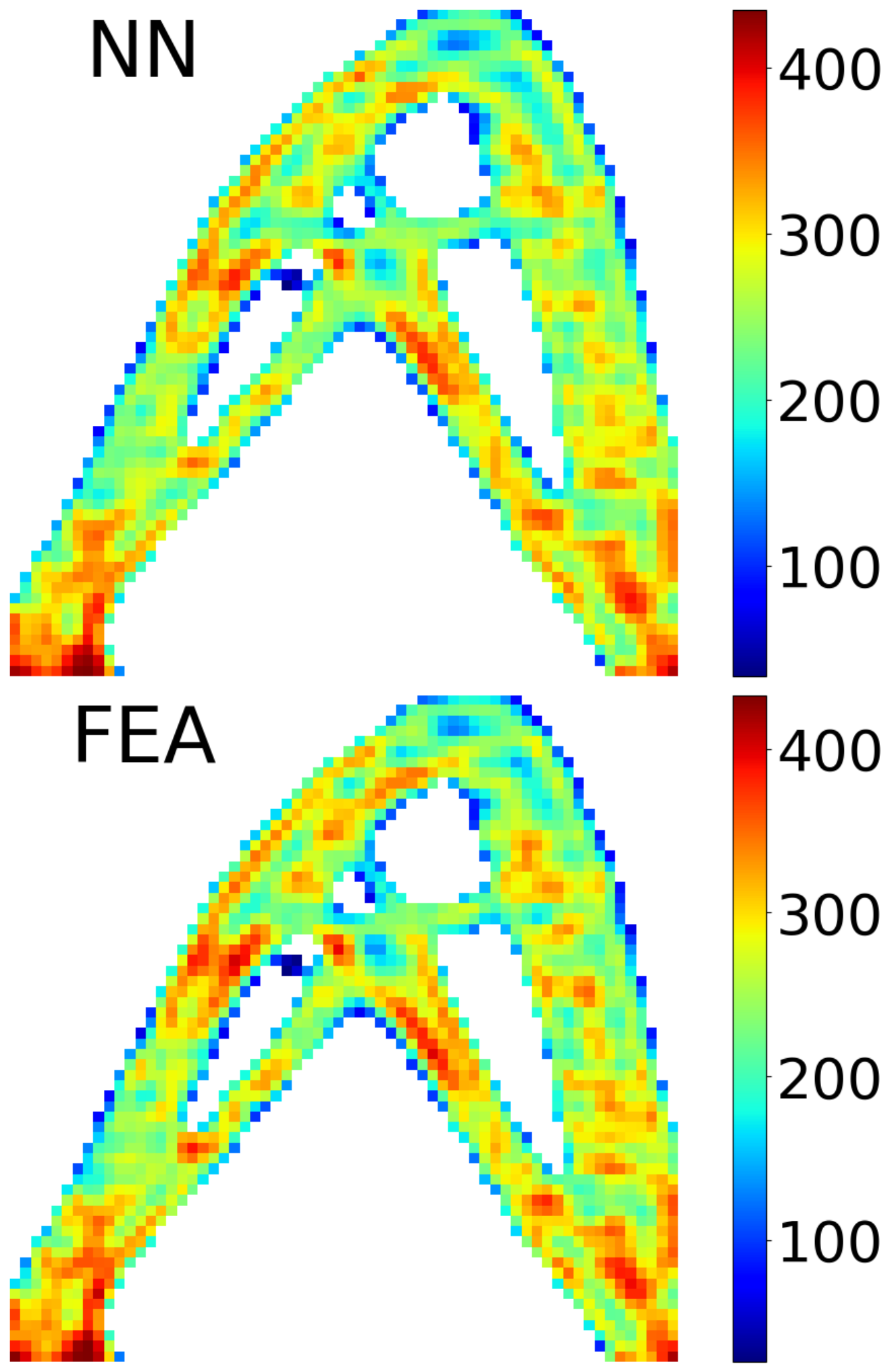}
         \label{rdon_stress_contour_75}
     }
     \subfloat[Worst case]{
         \includegraphics[trim={0cm 0cm 0cm 0cm},clip,width=0.20\textwidth]{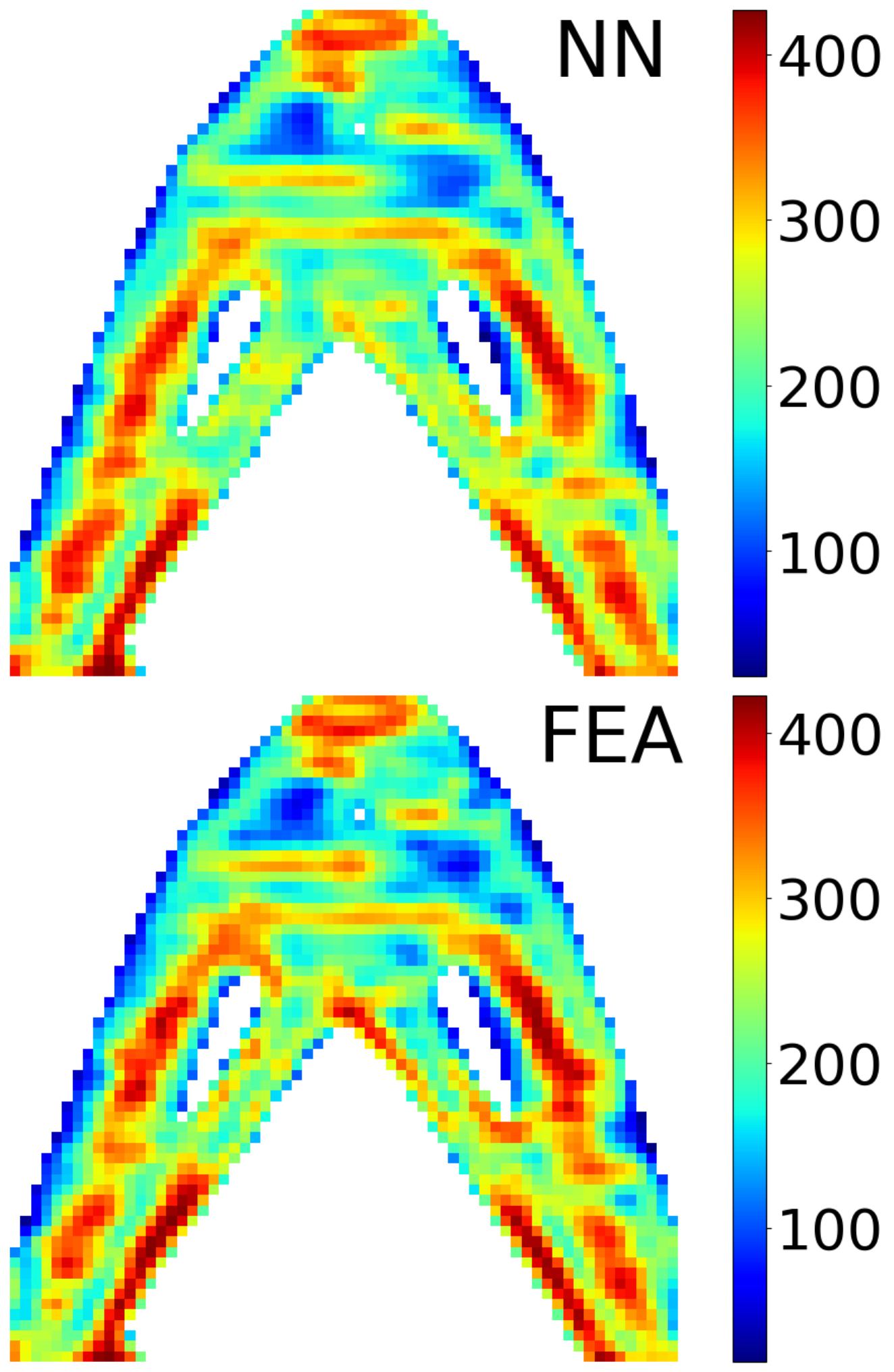}
         \label{rdon_stress_contour_100}
     }
    \caption{Stress predictions by ResUNet-based DeepONet compared to FEA for samples in \tref{rdon_stress_samples}. (Rel. $L_2$ Error)}
    \label{rdon_stress_contour}
\end{figure}

On the other hand, statistics of stress predictions are in \tref{rdon_stress_samples}. Similarly, the term "NN" refers to the stress field predictions obtained through a ResUNet-based DeepONet approach, while "FEA" denotes the results derived from finite element analysis. In additive manufacturing processes, the temperature will be cooled down in a product when in use. However, the residual stress remains in the sample. The coupled multiphysics nature with temperature makes the residual stress distribution chaotic, as illustrated in \fref{rdon_stress_contour}. ResUNet-based DeepONet was successful in predicting such convoluted distribution data, having the worst case relative $L_2$ error 9.27\%.

\begin{table}[h!]
    \caption{Error statistics for von Mises stress}
    \small
    \centering
    \begin{tabular}{ccccccc}
      & \vline 
      & {\textbf{Best case}} 
      & {\textbf{25$^{th}$ Percentile}}
      & {\textbf{50$^{th}$ Percentile}}
      & {\textbf{75$^{th}$ Percentile}}
      & {\textbf{Worst case}} \\
    \hline
    Rel. $L_2$ error (\textbf{\%}) 
    & \vline  
    & 3.28 
    & 4.79 
    & 5.37 
    & 6.16 
    & 10.2\\
    
    MAE (\textbf{MPa}) 
    & \vline  
    & 4.41 
    & 4.71 
    & 6.32 
    & 5.53 
    & 10.6\\
    
    \end{tabular}
    \label{rdon_stress_samples}
\end{table}

Reviewing all stress distributions, the Mean Relative L2 Error (MRL2E) stood at 5.53\%, the MRAE at 4.43\%, the average MAE at 5.70 MPa, and the CoP was determined to be 0.994 for von Mises stress predictions. The performance metrics of the model regarding von Mises stress predictions across all test samples are compiled in Table \ref{rdon_results}.

\begin{table}[h!]
    \caption{Statistics of ResUNet-based DeepONet performance}
    \small
    \centering
    \begin{tabular}{cccccc}
      & \vline 
      & \multicolumn{1}{c}{\textbf{MRL2E}} 
      & \multicolumn{1}{c}{\textbf{MRAE}} 
      & \multicolumn{1}{c}{\textbf{Average MAE}} 
      & \multicolumn{1}{c}{\textbf{CoP}} \\
      
    \hline
    Stress 
    & \vline  
    & 5.53\% 
    & 4.43\% 
    & 5.70MPa 
    & 0.994\\
    
    Temperature 
    & \vline  
    & 3.14\% 
    & 2.31\% 
    & 6.37$^{\circ}C$ 
    & 0.998\\
    
    \end{tabular}
    \label{rdon_results}
\end{table}

\begin{figure}[h!] 
    \centering
     \subfloat[Temperature Rel. $L_2$ error histogram]{
         \includegraphics[trim={0cm 0cm 0cm 0cm},clip,width=0.5\textwidth]{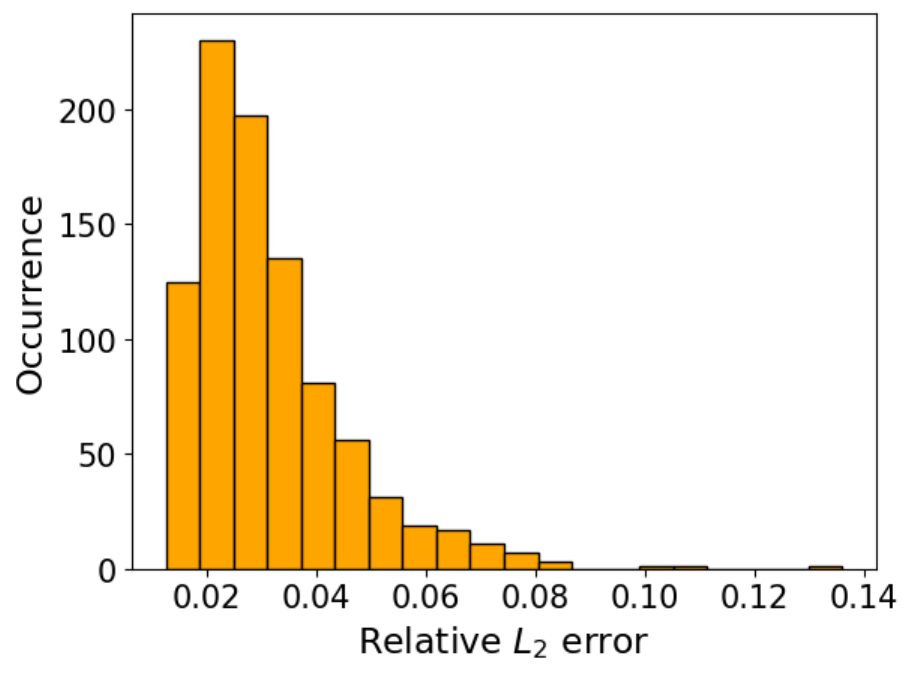}
         \label{rdon_stress_hist}
     }
     \subfloat[von Mises stress Rel. $L_2$ error histogram]{
         \includegraphics[trim={0cm 0cm 0cm 0cm},clip,width=0.5\textwidth]{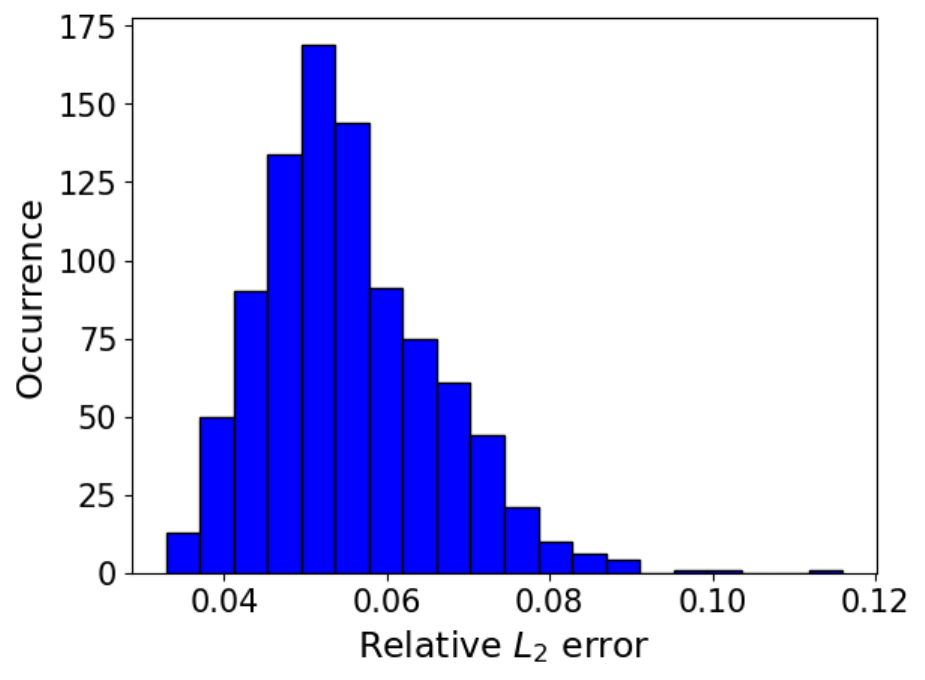}
         \label{rdon_temp_hist}
     }
    \caption{Relative $L_2$ error histograms for test sampels}
    \label{rdon_hist}
\end{figure}

\fref{rdon_hist} presents a histogram of the Relative $L_2$ errors for the test samples, encompassing both temperature and stress predictions. The distributions in both cases exhibit left-skewed statistics, clustering closely around the mean value of each plot.

\begin{figure}[h!] 
    \centering
         \includegraphics[width=0.6\textwidth]{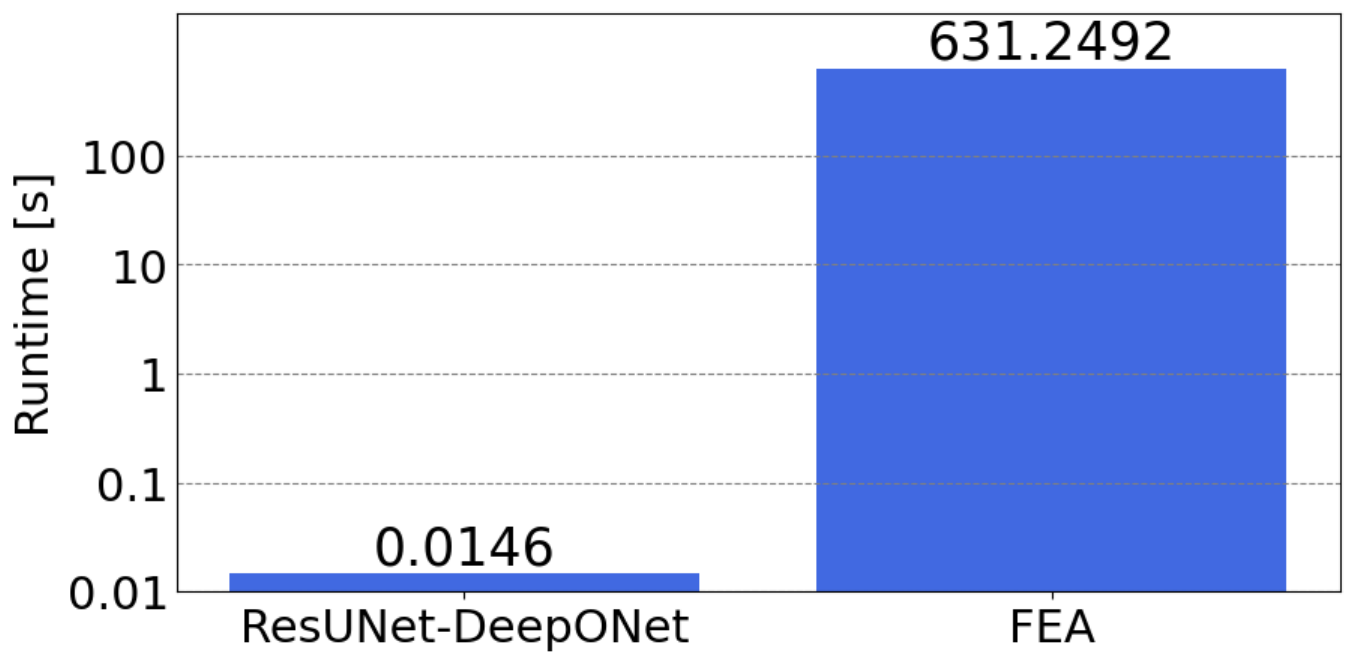}
    \caption{Runtime Comparison Between FEA and ResUNet-based DeepONet per one sample}
    \label{rdon_speedup}
\end{figure}

Lastly, the inference time for test samples is 13.4 seconds for 915 designs, making one sample take 0.0146 seconds to predict. To compare, running FEA simulations for one design takes 631 seconds on average, having the neural network prediction 4.3 x10$^4$ times faster. The comparison is illustrated in \fref{rdon_speedup}. This enhancement in efficiency underscores the neural network's ability to undergo swift assessments, reducing the computational time in huge scales necessary for predicting multiphysics properties in real time.



\subsection{Insights from the AM process using trained DeepONet model}
\label{sec:insights}

Understanding and mitigating residual stresses during the AM process is crucial for ensuring printed components' structural integrity and performance. This simple application example illustrates the enormous computational efficiency of the trained ResUNet-based DeepONet model defined in \sref{sec:resunet_and_deeponet} to swiftly predict residual stress in unseen 2D TO designs \citep{HE2023116277}, based on different printing speeds, as detailed in \sref{sec:AM_data_generation}. For this purpose, the trained network was used to predict the residual stress field for 915 new design sets incorporating different printing speeds, which were previously unseen by the network. From the predictions, the maximum residual stress for each design was extracted. The trained model finished all evaluations in 2.27 seconds, which corresponds to only $2.49 \times 10^{-3}$ seconds per case. It would take many days on a high-end workstation to calculate the stress field for these 915 new designs, even with a highly optimized FEA code with a parallel solver. 

\begin{figure}[h!] 
    \centering
     \subfloat[Maximum stress comparison for two different \\ 
                printing speeds, 7.5mm/s and 17.5mm/s on neural \\ 
                network predictions]{
         \includegraphics[trim={0cm 0cm 0cm 0cm},clip,width=0.5\textwidth]{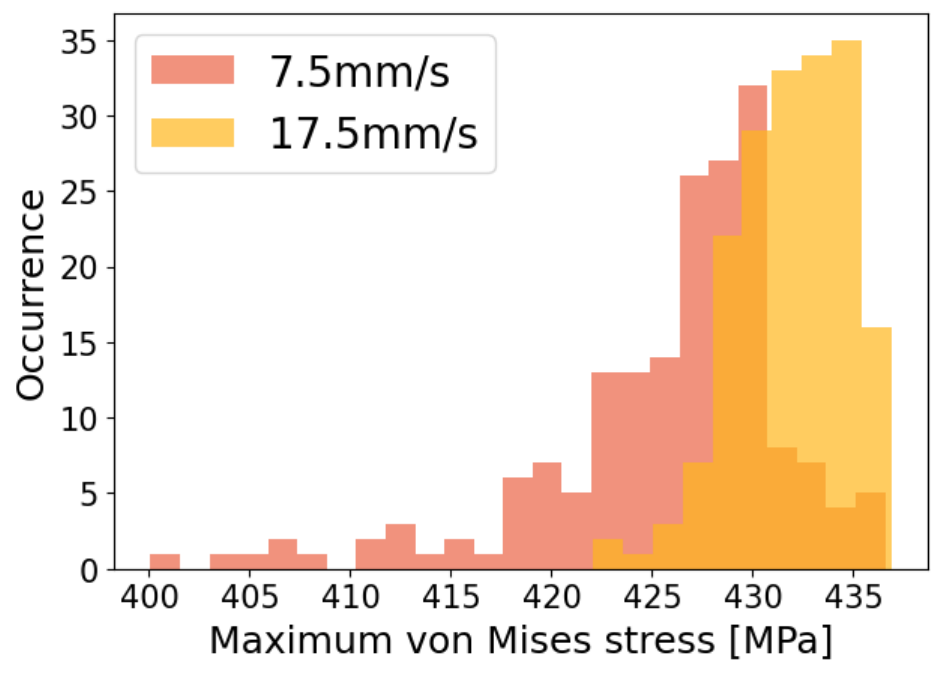}
         \label{Max_diff_des}
     }
     \subfloat[Averaged maximum von Mises stress by printing velocity]{
         \includegraphics[trim={0cm 0cm 0cm 0cm},clip,width=0.5\textwidth]{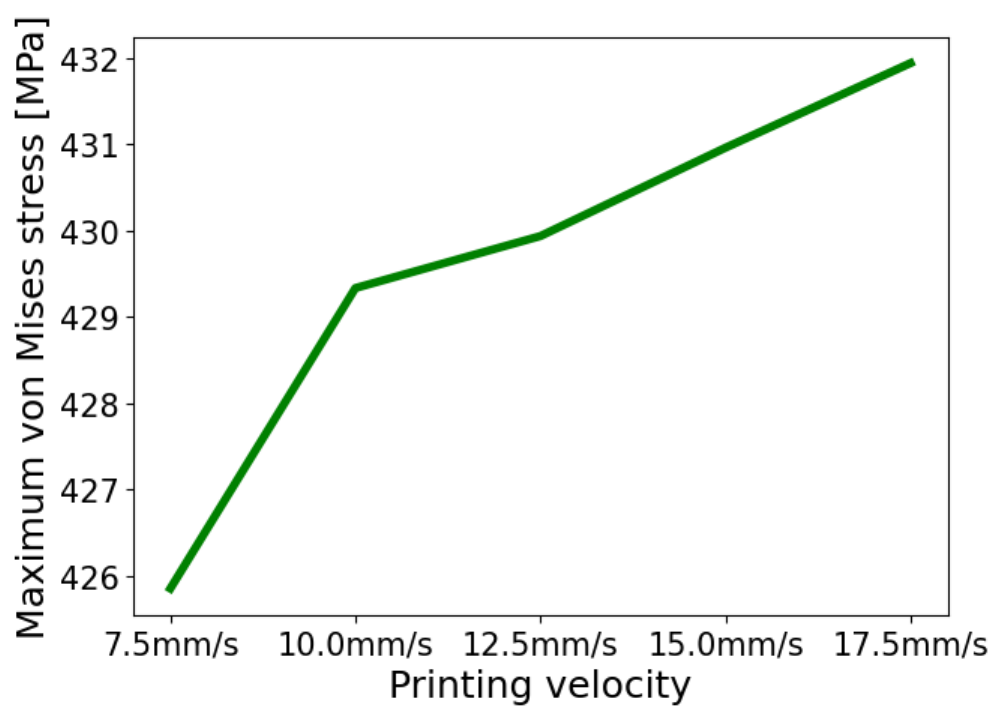}
         \label{Max_stress_by_speed}
     }
    \caption{Correlation between maximum von Mises stress and printing velocity}
    \label{Max_stress_by_vel}
\end{figure}

Two printing velocities of 7.5mm/s and 17.5mm/s were compared in \fref{Max_diff_des} to determine the maximum stress variation over different designs. It could be inferred that the maximum stress depends on both printing speed in the AM process and the underlying design geometries. Samples produced at a printing speed of 17.5mm/s exhibited a higher incidence of high-stress values (larger than 430 Mpa) in comparison to those fabricated at a speed of 7.5mm/s. \fref{Max_stress_by_speed} shows the relationship between the average max stress (across all 915 samples) and printing speeds. It can be seen that when the printing velocity increases, the average maximum residual stress generally increases. 

\begin{figure}[h!] 
    \centering
     \subfloat[maximum stress design]{
         \includegraphics[trim={0cm 0cm 0cm 0cm},clip,width=0.23\textwidth]{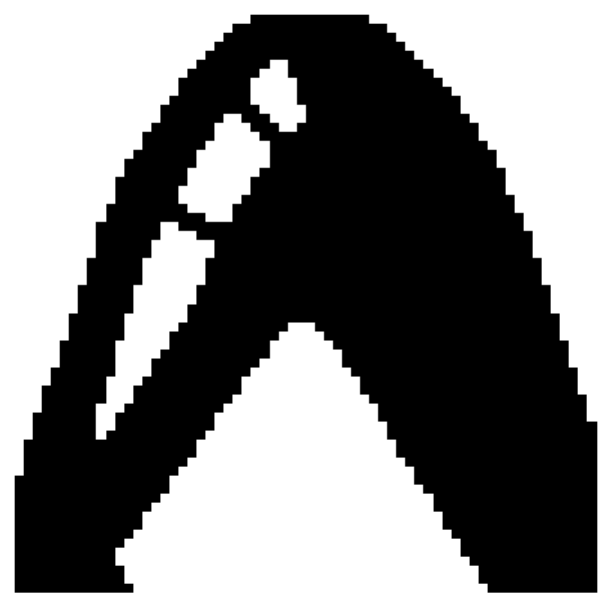}
         \label{Max_des}
     }
     \subfloat[Median stress design]{
         \includegraphics[trim={0cm 0cm 0cm 0cm},clip,width=0.23\textwidth]{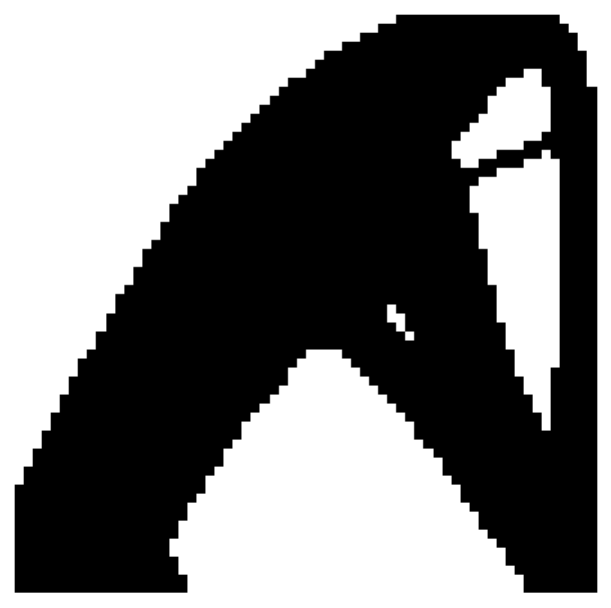}
         \label{Med_des}
     }
     \subfloat[Minimum stress design]{
         \includegraphics[trim={0cm 0cm 0cm 0cm},clip,width=0.23\textwidth]{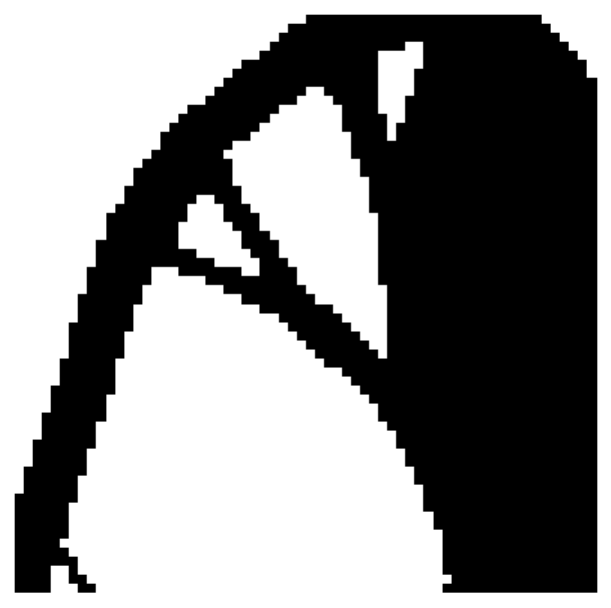}
         \label{Min_des}
     }
    \caption{Three designs chosen from the scalar \textit{maximum-stress values}}
    \label{Max_min_med_des}
\end{figure}

\begin{figure}[h!] 
    \centering
         \includegraphics[width=0.8\textwidth]{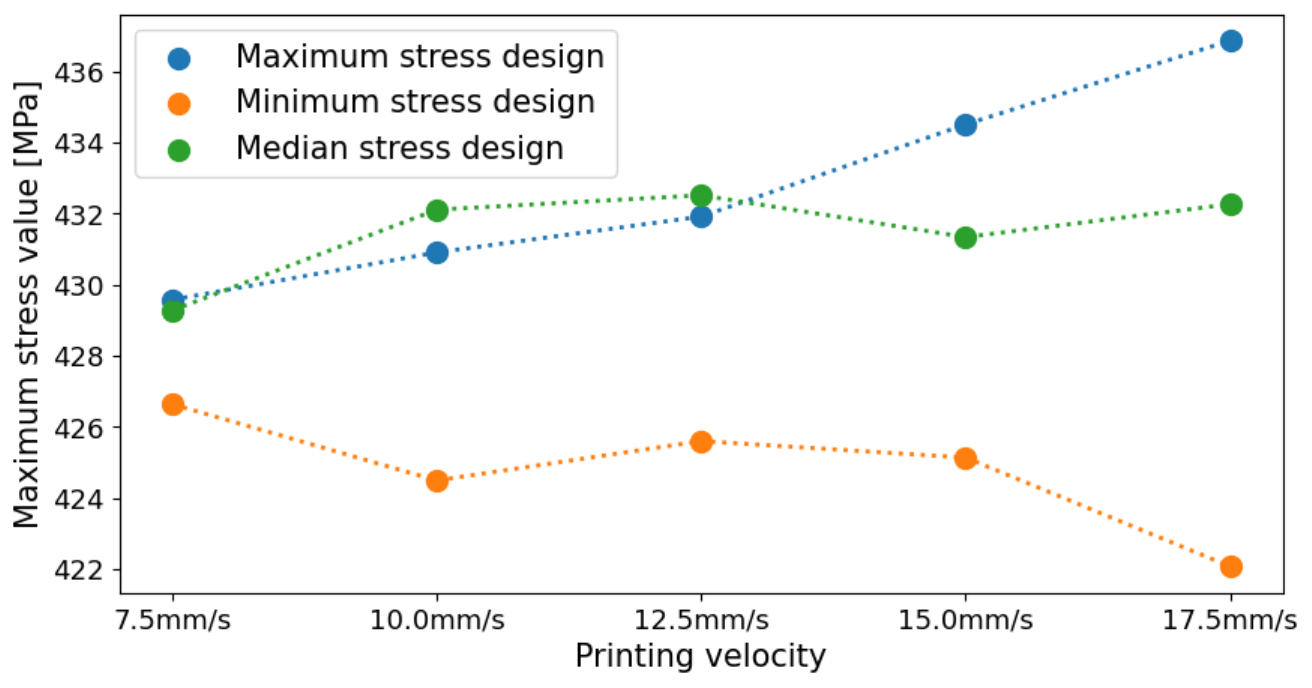}
    \caption{Residual stress values of three designs}
    \label{Max_min_med_graph}
\end{figure}

Three designs were selected to further investigate the influence of the printing speed on the residual stress of individual designs. From the scalar maximum residual stress representing every sample predicted by the network, the maximum, median, and minimum values of the maximum residual stress were chosen based on 17.5 mm/s printing velocity. This particular velocity was chosen based on observed data in \fref{Max_stress_by_vel}, which indicated an increasing trend of maximum residual stress. The strategy was aimed at incorporating and examining the design responsible for the highest overall residual stress.
Therefore, 17.5 mm/s, the highest printing velocity, was selected as it resulted in a higher average maximum stress compared to other velocities. \fref{Max_min_med_des} provides their geometry representation. \fref{Max_min_med_graph} delineates how the maximum stress in each design behaves as printing velocity varies. Although the average maximum stress increases with the printing velocity for this dataset, possessing 915 designs in total, as shown in \fref{Max_stress_by_speed}, the individual designs can show variable trends for the printing velocity (see \fref{Max_min_med_graph}). The predictions from the ResUNet-based DeepONet model offer quick insights into feasible printing speeds for specific designs and can generally serve as a very effective tool for preliminary designs and optimizations of AM processes. 

\section{Conclusions and future work}
\label{sec:conc}

This novel work demonstrates the capabilities of advanced and very recently devised variants of DeepONet architecture trained with data from multiphysics simulations in materials processing and additive manufacturing to infer entire thermal and mechanical solution fields almost instantly for unknown parametric inputs that include loads, loading histories, process parameters, and even variable geometries.  Even though the application areas are different, representing an established materials processing method for steel making and LDED AM process, in both cases, DeepONets were able to provide relatively accurate solutions in the entire domains, which makes this methodology potentially applicable across a wide range of classical and advanced manufacturing processes.  

Only the data from the final time step was used for training and predicting final solution fields. In future studies, every simulation time step will be incorporated into neural networks to predict the entire history of solution fields. Additionally, the exploration of additive manufacturing for intricate 3D structures will be studied to predict mechanical properties, extending the DeepONet architecture to handle 3D geometries. The data in this work was produced from macro-level multiphysics simulations and generated relatively quickly on a high-performance computing (HPC) system within a couple of days. In our future works, we will use even more high-throughput and parallel capabilities of HPC to generate data from existing multi-scale computational frames, including information from some vital micro and mesoscale phenomena and more process parameters. Since DeepONets do not have to be re-trained with new parametric inputs, their data has to be generated only once. Moreover, the trained DeepONet models can be transferred to laptops and desktops to infer results quickly without any modeling software or HPC.  

In conclusion, this research, for the first time, proves that adequately trained advanced DeepONet architectures can serve as a surrogate for modeling complex real-world manufacturing processes and thus paves the way for future preliminary optimization, inverse and sensitivity analysis, uncertainty quantification, online controls, digital twins and similar iterative numerical analysis that require computationally challenging and expensive multiphysics forward solution evaluations in entire domains with variable geometries, loads, and other input process parameters.

\section*{Replication of results}
The data and source code supporting this study will be available in a public GitHub repository.

\section*{Conflict of interest}
The authors declare that they have no conflict of interest.

\section*{Acknowledgements}
The authors would like to thank the National Center for Supercomputing Applications (NCSA) at the University of Illinois, and particularly its Research Computing Directorate, the Industry Program, and the Center for Artificial Intelligence Innovation (CAII) for their support and hardware resources. This research is a part of the Delta research computing project, which is supported by the National Science Foundation (award OCI 2005572) and the State of Illinois, as well as the Illinois Computes program supported by the University of Illinois Urbana-Champaign and the University of Illinois System.

\section*{CRediT author contributions}
\textbf{Shashank Kushwaha}: Conceptualization, Methodology, Software, Formal analysis, Investigation, Writing - Original Draft.
\textbf{Jaewan Park}: Conceptualization, Methodology, Software, Formal analysis, Investigation, Writing - Original Draft.
\textbf{Seid Koric}: Conceptualization, Methodology, Supervision, Resources, Writing - Original Draft, Funding Acquisition.
\textbf{Junyan He}: Investigation, Writing - Review \& Editing. 
\textbf{Iwona Jasiuk}: Supervision, Writing - Review \& Editing.
\textbf{Diab Abueidda}: Conceptualization, Methodology, Supervision, Writing - Review \& Editing.

\bibliographystyle{unsrtnat}
\setlength{\bibsep}{0.0pt}
{\scriptsize \bibliography{References.bib} }

\end{document}